\documentclass[a4paper,11pt]{article}
\usepackage{jcappub} % for details on the use of the package, please see the JINST-author-manual
\usepackage{lineno}
\usepackage{graphicx}
\usepackage{dcolumn}
\usepackage{bm}
\usepackage{amssymb}
\usepackage{latexsym}
\usepackage{booktabs}
\usepackage{amsmath}
\usepackage{multirow}
\usepackage{url}
\usepackage{footnote}
\usepackage{float}
\usepackage{acro}
\usepackage{hyperref}
\usepackage{cleveref}
\usepackage[section]{placeins}
\DeclareAcronym{PBH}{
  short = PBH ,
  long  = primordial black hole ,
  short-plural = s ,
}

\DeclareAcronym{CMB}{
  short = CMB ,
  long  = cosmic microwave background ,
  short-plural =  ,
}

\DeclareAcronym{LSS}{
  short = LSS ,
  long  = large-scale structures ,
  short-plural =  ,
}

\DeclareAcronym{IGM}{
  short = IGM ,
  long  = intergalactic medium ,
  short-plural =  ,
}

\DeclareAcronym{GW}{
  short = GW ,
  long  = gravitational wave ,
  short-plural = s ,
}

\DeclareAcronym{MCGs}{
  short = MCGs ,
  long  = molecular-cooling galaxies ,
  short-plural =  ,
}

\DeclareAcronym{ACGs}{
  short = ACGs ,
  long  = atom-cooling galaxies ,
  short-plural =  ,
}

\DeclareAcronym{EoR}{
  short = EoR ,
  long  = epoch of reionization ,
  short-plural =  ,
}

\DeclareAcronym{EDGES}{
  short = EDGES ,
  long  = Experiment to Detect the Global Epoch of Reionization Signature ,
  short-plural =  ,
}

\DeclareAcronym{Fermi-LAT}{
  short = Fermi-LAT ,
  long  = Fermi Large Area Telescope ,
  short-plural =  ,
}

\DeclareAcronym{INTEGRAL}{
  short = INTEGRAL ,
  long  = International Gamma-Ray Astrophysics Laboratory,
  short-plural =  ,
}

\DeclareAcronym{SARAS 3}{
  short = SARAS 3,
  long  = Shaped Antenna measurement of the background Radio Spectrum 3 ,
  short-plural =  ,
}

\DeclareAcronym{DM}{
   short = DM ,
   long  = dark matter ,
   short-plural =  ,
}

\DeclareAcronym{HEAO}{
   short = HEAO,
   long  = High Energy Astrophysical Observatory ,
   short-plural =  ,
}

\DeclareAcronym{COMPTEL}{
   short = COMPTEL ,
   long  = Imaging Compton Telescope ,
   short-plural =  ,
}

\DeclareAcronym{EGRET}{
   short = EGRET ,
   long  = Energetic Gamma-ray Experiment Telescope ,
   short-plural =  ,
}

\DeclareAcronym{VERITAS}{
   short = VERITAS ,
   long  = Very Energetic Radiation Imaging Telescope Array System ,
   short-plural =  ,
}

\DeclareAcronym{H.E.S.S}{
   short = H.E.S.S,
   long  = High Energy Stereoscopic System ,
   short-plural =  ,
}

\DeclareAcronym{MAGIC}{
   short = MAGIC,
   long  = Major Atmospheric Gamma Imaging Cherenkov telescopes ,
   short-plural =  ,
}

\DeclareAcronym{LIGO}{
   short = LIGO,
   long  = Laser Interferometer Gravitational-Wave Observatory ,
   short-plural =  ,
}

\DeclareAcronym{VIRGO}{
   short = VIRGO,
   long  = Virgo Gravitational Wave Interferometer  ,
   short-plural =  ,
}

\def\blue#1{{\textcolor{blue}{#1}}}

\title{Prospects for probing dark matter particles and primordial black holes with the Hongmeng mission using the 21 cm global spectrum at cosmic dawn}

\author[a]{Meng-Lin Zhao,}
\author[b,\ast]{Sai Wang\note[$\ast$]{Corresponding author.}}
\author[a,c,d,\ast]{and Xin Zhang}

\affiliation[a]{Liaoning Key Laboratory of Cosmology and Astrophysics, College of Sciences, Northeastern University, Shenyang 110819, China}
\affiliation[b]{School of Physics, Hangzhou Normal University, Hangzhou 311121, China}
\affiliation[c]{National Frontiers Science Center for Industrial Intelligence and Systems Optimization, Northeastern University, Shenyang 110819, China}
\affiliation[d]{MOE Key Laboratory of Data Analytics and Optimization for Smart Industry, Northeastern University, Shenyang 110819, China}

\emailAdd{zhaoml@stumail.neu.edu.cn}
\emailAdd{wangsai@hznu.edu.cn}
\emailAdd{zhangxin@mail.neu.edu.cn}

\abstract{
Probing dark matter particles and primordial black holes remains a pivotal challenge in modern cosmology. Exotic energy injections from dark matter annihilation, decay, and PBH Hawking evaporation can alter the thermal and ionization histories of the early universe, leaving distinctive imprints on the 21 cm global spectrum. We assess the potential of the upcoming space project, the Hongmeng mission, to probe dark matter particles and PBHs using the 21 cm global spectrum. Under ideal conditions with 1000 hours of integration time and negligible foreground residuals, the Hongmeng project can reach  sensitivities to dark matter annihilation cross sections and decay lifetimes to $\langle \sigma v \rangle \sim 10^{-28}\,\mathrm{cm^3\,s^{-1}}$ and $\tau \sim 10^{28}\,\mathrm{s}$, respectively, for dark matter particles with a mass of $10\,\mathrm{GeV}$. It can also probe PBHs with masses of $10^{16}\,\mathrm{g}$ and abundances as low as $f_{\mathrm{PBH}} \simeq 10^{-6}$. These results indicate that the Hongmeng mission can improve current constraints on dark matter annihilation, decay, and PBH Hawking radiation by nearly two orders of magnitude. Moreover, the Hongmeng mission surpasses current limits on sub-GeV dark matter probing and enables the probing of Hawking radiation from PBHs with masses above $10^{17}\,\mathrm{g}$, which remain undetectable through conventional cosmological means. Overall, the upcoming Hongmeng project holds great promise for advancing the investigation of both dark matter and PBHs, potentially deepening our understanding of the nature of dark matter.

}

\begin{document}

\maketitle

\flushbottom
\section{Introduction}
\label{sec:INTRODUCTION}

The concept of \ac{DM} is postulated to account for a significant portion of the universe's mass, as evidenced by observations of galactic rotations, galaxy cluster dynamics, gravitational lensing, the anisotropies in the \ac{CMB}, and other phenomena (see reviews in Refs.~\cite{ParticleDataGroup:2024cfk,Bechtol:2019acd,Schumann:2019eaa,Bertone:2016nfn,Gaskins:2016cha,Feng:2010gw,Bertone:2004pz,Sahni:2004ai,Smith:1988kw} and references therein). Despite being invisible, the gravitational influence of \ac{DM} plays a crucial role in the formation and evolution of cosmic structures. As a result, a variety of candidate models for \ac{DM} have been proposed in the literature \cite{Lin:2019uvt,Feng:2010gw,Bertone:2004pz}. These candidate models can generally be classified into three categories: particles, macroscopic objects, and modifications of gravity, each encompassing a diverse array of theoretical frameworks. In this study, our focus is primarily on the former two categories.

The annihilation and decay of \ac{DM} particles are pivotal processes within the domain of \ac{DM} research (see reviews in Refs.~\cite{Fermi-LAT:2016uux,MAGIC:2016xys,Slatyer:2016qyl,Fermi-LAT:2015att,Slatyer:2015jla,HESS:2013rld,Steigman:2012nb,Chen:2003gz,Gondolo:1999ef,Cheung:2018vww}). On one hand, \ac{DM} annihilation occurs when two \ac{DM} particles collide, resulting in the conversion to other particles and the potential generation of detectable signals such as gamma rays or cosmic rays. These signals offer valuable insights into the interactions among \ac{DM} particles. On the other hand, \ac{DM} decay involves the spontaneous conversion of a \ac{DM} particle into lighter particles, providing essential information on the stability and lifetime of \ac{DM} particles. Investigations into these processes are crucial for advancing our understanding of \ac{DM} and present unique opportunities to probe the nature of \ac{DM} through astronomical observations.

Should \acp{PBH} constitute a significant portion of \ac{DM}, the observation and analysis of their Hawking radiation emissions hold promise for elucidating their fundamental characteristics, e.g., their abundance as a function of their masses. The theoretical framework of Hawking radiation, as proposed by Stephen Hawking \cite{Hawking:1971ei}, posits that black holes can emit radiation and particles such as gamma rays, cosmic rays, and neutrinos, due to quantum effects near the event horizon. Through the meticulous study of this radiation, we can not only deepen our understanding of \acp{PBH} as a viable \ac{DM} candidate, but also find potential implications for the nature of quantum gravity. 
Relevant constraints on the \ac{PBH} mass function are shown in Refs.~\cite{Sasaki:2018dmp,Carr:2020gox,Green:2020jor, Carr:2021bzv,Auffinger:2022ive,Mittal:2021egv}. 

The 21 cm signal plays a pivotal role as an essential tool in the study of \ac{DM} \cite{Mena:2019nhm,Saha:2021pqf,Facchinetti:2023slb,Shao:2023agv,Novosyadlyj:2024bie}. Processes involving \ac{DM} particle annihilation and decay, as well as \ac{PBH} Hawking radiation emissions, can introduce exotic energy into the \ac{IGM} \cite{Sun:2023acy,Liu:2023fgu,Liu:2018uzy}, leading to significant alterations in the thermal and ionization histories of the universe. These modifications impact the spin temperature of neutral hydrogen, thereby influencing the 21 cm brightness temperature. The importance of the 21 cm signal in unraveling the mysteries of \ac{DM} is underscored by these connections. Notably, compared to other cosmological probes such as the \ac{CMB} \cite{Slatyer:2016qyl,Liu:2023nct,Xu:2024vdn,Capozzi:2024gqy} and Lyman-$\alpha$ forest \cite{Liu:2020wqz}, the 21 cm absorption lines demonstrate enhanced sensitivity to phenomena occurring during the era of cosmic dawn with redshifts $z\sim10-20$. 
In particular, the \ac{EDGES} reported an absorption feature in the 21 cm global signal at $z\simeq17$ \cite{Bowman:2018yin}. 
Analyzing this signal, researchers placed upper limits on parameters characterizing the \ac{DM} particle annihilation and decay \cite{Liu:2018uzy} as well as the \ac{PBH} mass function \cite{Saha:2021pqf,Cang:2021owu}. 
However, such a signal was recently contradicted by the \ac{SARAS 3} at 95.3\% confidence level \cite{Bevins:2022ajf}.

%全称
Looking ahead, the Hongmeng project (also known as the Discovering the Sky at the Longest
Wavelength Mission) \cite{2023ChJSS..43...43C} offers significant advantages for 21 cm signal observations.
%鸿蒙阵列的介绍。
%鸿蒙绕月卫星阵列一共由十颗卫星组成，其中一颗用于星间通讯，有八颗低频卫星用于黑暗时代的成像观测，另有一颗高频卫星用于测量宇宙黎明时期的整体谱。
The Hongmeng project deploys a lunar-orbit interferometric array comprising one primary satellite and nine subsatellites to achieve full-sky observations across the $0.1-120$\,MHz frequency range.
The primary satellite is dedicated to satellite-to-satellite communication.
Among the subsatellites, eight perform interferometric measurements in the $0.1-30$\,MHz band to generate ultra-long wavelength sky maps.
The remaining subsatellite carries a high-frequency spectrometer dedicated to detecting the global spectrum in the $30-120$\,MHz band.
%相比于其他的地面的21厘米整体谱实验，鸿蒙卫星受地面杂波和大气层的影响较小，因此，预期鸿蒙能够得到一个更精确的关于21厘米整体谱的测量。
Compared to ground-based 21 cm global signal experiments, the Hongmeng project is less affected by environmental noise, thereby improving measurement precision.
The lunar-orbit observation by Hongmeng effectively avoids Earth's ionospheric distortion and terrestrial radio frequency interference, which enhances detection accuracy relative to ground-based observations.

In this study, we systematically evaluate the Hongmeng mission's capacity to probe signatures originating from \ac{DM} annihilation, decay, and \ac{PBH} Hawking radiation.
The subsequent sections of this paper are structured as follows. Section \ref{sec:Exotic energy injection} presents the scenarios of exotic energy injection considered in this study, with their effects on the \ac{IGM} and 21 cm global signal summarized in Section \ref{sec:Theory}. Section \ref{sec:Fisher} elaborates on the Fisher information matrix and parameter settings employed, while Section \ref{sec:RESULTS} showcases the primary findings. Finally, Section \ref{sec:CONCLUSION} offers a concise discussion and summary of the study's outcomes.
In our present study, we adhere to the cosmological parameters outlined in the Planck 2018 results \cite{Planck:2018vyg}, specifically utilizing $(\Omega_{\rm m},\Omega_{\rm b},\Omega_{\rm \Lambda},h,\sigma_{\rm 8},n_{\rm s}) = (0.31, 0.049, 0.69, 0.68, 0.81, 0.97)$.
Here $\Omega_{\rm m}$, $\Omega_{\rm b}$, and $\Omega_{\rm \Lambda}$ represent the present-day energy-density fractions of non-relativistic matter, baryons, and dark energy, respectively, while $h$ is the dimensionless Hubble constant, $\sigma_{\rm 8}$ is the amplitude of matter fluctuations, and $n_{\rm s}$ is the spectral index of primordial curvature perturbations. Throughout this work, the speed of light is denoted as $c$, and the Boltzmann constant as $k_{\rm B}$.

\section{Scenarios of exotic energy}
\label{sec:Exotic energy injection}

In this section, we briefly summarize the formulas for scenarios of exotic energy considered in our present work. 
The present-day energy density of \ac{DM} is expressed as $\rho_{\rm DM}=\rho_{c}\Omega_{\rm DM}$, where $\Omega_{\rm DM}=\Omega_{\rm m}-\Omega_{\rm b}$, and $\rho_{c}$ denotes the critical energy density of the present-day cosmos. 

\subsection{Annihilation and decay of DM particles}

The annihilation and decay of \ac{DM} particles can result in the production of particles from the standard model, encompassing both primary and secondary products.
We consider three categories of primary particles: photons, electron-positron pairs, and bottom-anti-bottom quark pairs.
Secondary particles, such as photons, electrons, protons, neutrinos, and others, are generated from primary particles through various processes including annihilation, decay, and hadronization \cite{Slatyer:2015kla}. 
To simulate these processes, we use the \texttt{PPPC4DMID} \cite{Cirelli:2010xx} and \texttt{pythia} \cite{Bierlich:2022pfr}.
In the subsequent section, our focus is specifically on photons, electrons, and positrons due to their efficiency in depositing energy into the \ac{IGM} \cite{Slatyer:2012yq,Slatyer:2009yq}.

We consider an s-wave annihilation channel of \ac{DM} particles, without specifying the primary products. 
The exotic energy injected into the \ac{IGM} per unit volume per unit time is given by \cite{Liu:2019bbm}
\begin{equation}
    \left(\frac{{\rm d}E}{{\rm d}V{\rm d}t}\right)_{\rm inj,ann} = \rho^{2}_{\rm DM} \mathcal{B}(z) (1+z)^{6} c^2 \frac{\langle \sigma v \rangle}{m_{\chi}}\ , \label{eq:darkmatterani}
\end{equation}
where $\mathcal{B}(z)$ represents a boost factor accounting for the clumping of dark matter, which has been extensively studied in Ref.~\cite{Takahashi:2021pse}, $z$ denotes the cosmological redshift, $\langle \sigma v \rangle$ is the thermally-averaged annihilation cross-section of dark matter particles, and $m_{\chi}$ represents the mass of \ac{DM} particles. 
In this context, we assume that all \ac{DM} particles are capable of annihilation, implying $f_{\rm ann}=1$.
Otherwise, a factor of $f_{\rm ann}^{2}$ should be included on the right-hand side of the above formula.

The exotic energy injected into the \ac{IGM} per unit volume per unit time from \ac{DM} particle decay, regardless of the primary products, is given by \cite{Liu:2019bbm}
\begin{equation}
    \left(\frac{{\rm d}E}{{\rm d}V{\rm d}t}\right)_{\rm inj,dec} = \rho_{\rm DM} (1+z)^{3} c^2 \frac{1}{\tau}\ , \label{eq:darkmatterdecay} 
\end{equation}
where $\tau$ represents the lifetime of the dark matter particles. We assume that all dark matter particles can decay, i.e., $f_{\rm dec} = 1$.
If this is not the case, a factor of $f_{\rm dec}$ should be included on the right-hand side of the above formula.

%For a decay channel of \ac{DM} particles, regardless of primary products, we get the exotic energy injected into \ac{IGM} gas per unit volume per unit time of the form \cite{Liu:2019bbm}
%\begin{equation}
%    \left(\frac{{\rm d}E}{{\rm d}V{\rm d}t}\right)_{\rm inj,dec} = %\rho_{\rm DM} (1+z)^{3} c^2 \frac{1}{\tau}\ , %\label{eq:darkmatterdecay} 
%\end{equation}
%where $\tau$ stands for a lifetime of \ac{DM} particles. 
%Here, we assume the fraction of \ac{DM} particles which can decay to be unity, i.e., $f_{\rm dec}=1$. Otherwise, we should multiply a factor of $f_{\rm dec}$ at the right hand side of the above formula. 

\subsection{Hawking radiation of PBHs}

The Hawking radiation of \acp{PBH} can produce particles of the standard model \cite{Hawking:1971ei}.
We adopt the \texttt{BlackHawk} \cite{Auffinger:2020ztk} to compute the particle spectra, denoted as ${\rm d}^{2} N / ({\rm d} E {\rm d} t)$. 
For \acp{PBH} in the mass range of $M_{\rm PBH} \sim 10^{15}-10^{18}$\,g, we focus on photons and electron-positron pairs as the dominant emission products. 
Therefore, the exotic energy injected into \ac{IGM} per unit volume per unit time is given by \cite{Slatyer:2015kla,Liu:2019bbm,Saha:2021pqf,Facchinetti:2023slb}
\begin{equation}
    \left(\frac{{\rm d}E}{{\rm d}V{\rm d}t}\right)_{\rm inj,PBH} = \int_{0}^{5\rm GeV} \frac{{\rm d}^{2} N}{{\rm d} E {\rm d} t} \Big|_{\rm \gamma} n_{\rm PBH} E {\rm d} E + \int_{m_{e}c^{2}}^{5\rm GeV} \frac{{\rm d}^{2} N}{{\rm d} E {\rm d} t}\Big|_{\rm e^{\pm}} n_{\rm PBH} (E - m_{\rm e} c^{2}) {\rm d} E\ , \label{eq:Einjtotal}
\end{equation}
where the first and second terms at the right hand side stand for contributions from photons and electron-positron pairs, respectively. 
Here, $m_{e}$ is the mass of electrons. 
$n_{\rm PBH}$ denotes the comoving number density of \acp{PBH}, given by \cite{Mena:2019nhm,Poulin:2017bwe}
\begin{equation}
    n_{\rm PBH} = \frac{f_{\rm PBH} \rho_{\rm DM} }{M_{\rm PBH}}\ , \label{eq:npbh}
\end{equation}
where $f_{\rm PBH}$ is the abundance of \acp{PBH} as \ac{DM}. 
In this work, we assume a monochromatic \ac{PBH} mass function for illustrative purposes.
Nevertheless, our research can be readily  extended to accommodate other mass functions of \acp{PBH} if necessary.

\section{Influence of exotic energy on the 21 cm global spectrum}
\label{sec:Theory}

Considering the exotic energy injection and deposition efficiency, we study the influence of the exotic energy on the thermal and ionization histories of the \ac{IGM}. 
Eventually, we demonstrate changes in the 21 cm global spectrum at cosmic dawn. 

The differential brightness temperature for the 21 cm global signal is defined as \cite{Furlanetto:2006jb,Pritchard:2011xb}
\begin{equation}
    {T_{\rm 21}(z)} = 23\, x_{\rm HI}(z) \left(\frac{0.15}{\Omega_{\rm m}}\right)^{\frac{1}{2}} \left(\frac{\Omega_{\rm b} h^{2}}{0.02}\right) \left(\frac{1+z}{10}\right)^{\frac{1}{2}} \left[1-\frac{T_{\rm CMB}(z)}{T_{\rm S}(z)}\right]~\rm{mK}\ , \label{eq:T21}
\end{equation}
where $x_{\rm HI}$ stands for the neutral fraction of hydrogen, $T_{\rm CMB}$ is the temperature of \ac{CMB} radiation at $z$, and $T_{\rm S}$ denotes the spin temperature of hydrogen at $z$. 
Defined as a ratio between the populations of triplets and singlet of neutral hydrogen, the spin temperature is given by \cite{Pritchard:2011xb}
\begin{equation}
    T_{\rm S}^{-1} = \frac{T_{\rm CMB}^{-1} + x_{\alpha}T_{\alpha}^{-1} + x_{\rm c}T_{\rm k}^{-1}}{1 + x_{\alpha} + x_{\rm c}}\ , \label{eq:TS}
\end{equation}
where $T_{\alpha}$ stands for the color temperature of the Lyman-$\alpha$ photons, $x_{\alpha}$ is a Lyman-${\alpha}$ coupling coefficient due to resonant scattering, i.e., the Wouthuysen-Field effect \cite{Furlanetto:2006jb,Facchinetti:2023slb}. $T_{\rm k}$ denotes the kinetic temperature of the \ac{IGM} gas, and $x_{\rm c}$ is a coupling coefficient due to collisions between two hydrogen atoms, hydrogen atoms and electrons, as well as hydrogen atoms and protons \cite{Furlanetto:2006jb,Pritchard:2011xb}.
Due to frequent scattering, the color temperature is tightly coupled to the kinetic temperature, i.e., $T_{\alpha}\simeq T_{\rm k}$. 
The exotic energy injection can change the evolution of $x_{\rm HI}$, $T_{\rm k}$, and $x_{\alpha}$, leading to changes in the 21 cm global signal. 

The evolution of $x_{\rm e}$ (i.e., the ionization fraction $x_{\rm e}=1-x_{\rm HI}$) and $T_{\rm k}$ is governed by the following system of equations \cite{Furlanetto:2006jb,Pritchard:2011xb,Mena:2019nhm,Facchinetti:2023slb}
\begin{eqnarray}
\frac{{\rm d} x_{\rm e}}{{\rm d} z} &=& \frac{{\rm d} t}{{\rm d} z} (\Lambda^{\rm exo}_{\rm ion} + \Lambda^{\rm X}_{\rm ion} + \alpha_{\rm A} C x^{2}_{\rm e} n_{\rm H})\ , \label{eq:xe} \\
\frac{{\rm d} T_{\rm k}}{{\rm d} z} &=&\frac{2}{3 k_{\rm B} (1+x_{\rm e})} \frac{{\rm d} t}{{\rm d} z} (\epsilon^{\rm exo}_{\rm heat}+\epsilon^{\rm X}_{\rm heat}+\epsilon^{\rm IC}_{\rm heat}) + \frac{2 T_{\rm k}}{3 n_{\rm b}} \frac{{\rm d} n_{\rm b}}{{\rm d} z} - \frac{T_{\rm k}}{1 + x_{\rm e}} \frac{{\rm d} x_{\rm e}}{{\rm d} z}\ . \label{eq:TK}
\end{eqnarray} 
Here, ${{\rm d}t}/{{\rm d}z} = {1}/[{H(z)(1+z)}]$, where $t$ denotes the cosmic time and $H(z)$ is the Hubble parameter at redshift $z$. 
$\alpha_{\rm A}$ stands for the case-A recombination coefficient. 
$C$ is the clumping factor. 
$n_{\rm H}$ and $n_{\rm b}$ denote the number densities of hydrogen and baryons, respectively. 
The ionizing rate per baryon due to astrophysical X rays is denoted as $\Lambda^{\rm X}_{\rm ion}$. 
The heating rates per baryon due to astrophysical X rays and inverse-Compton scattering are denoted as $\epsilon^{\rm X}_{\rm heat}$ and $\epsilon^{\rm IC}_{\rm heat}$, respectively. 
These astrophysical processes, as well as the coupled differential equations above, can be simulated using the \texttt{21cmFAST} package \cite{Mesinger:2010ne}. 

The ionization and heating rates per baryon due to exotic energy injection are given by  \cite{Mena:2019nhm,Facchinetti:2023slb}
\begin{eqnarray}
    \Lambda^{\rm exo}_{\rm ion} &=& F_{\rm HI}(z) \frac{1}{n_{\rm b}} \frac{n_{\rm H}}{n_{\rm b}} \frac{1}{E^{\rm HI}_{\rm ion}}  \left(\frac{{\rm d} E}{{\rm d} V {\rm d} t}\right)_{\rm inj, L} + F_{\rm He}(z)   \frac{1}{n_{\rm b}} \frac{n_{\rm He}}{n_{\rm b}} \frac{1}{E^{\rm He}_{\rm ion}} \left(\frac{{\rm d} E}{{\rm d} V {\rm d} t}\right)_{\rm inj, L}\ , \label{eq:lamdaexo}
    \\
    \epsilon^{\rm exo}_{\rm heat} &=& F_{\rm heat}(z) \frac{1}{n_{\rm b}} \left(\frac{{\rm d} E}{{\rm d} V {\rm d} t}\right)_{\rm inj, L}\ . \label{eq:epsilonheat}
\end{eqnarray}
Here, the subscript $_{\rm L}$ stands for annihilation, decay, and Hawking radiation, respectively. 
$n_{\rm He}$ is the number density of helium. 
$E^{\rm HI}_{\rm ion}$ and $E^{\rm He}_{\rm ion}$ denote the ionization energies of hydrogen and helium, respectively. 
In addition, $F_{\rm HI}$, $F_{\rm He}$, and $F_{\rm heat}$ denote for the energy deposition efficiencies for hydrogen ionization, helium ionization, and heating, respectively. 
These efficiencies can be calculated using \texttt{DarkHistory} \cite{Liu:2019bbm}. 

The Lyman-$\alpha$ coupling coefficient depends on the total flux of Lyman-$\alpha$ photons, i.e., \cite{Facchinetti:2023slb}
\begin{equation}
    x_{\alpha} = \frac{1.7 \times 10^{11}}{1 + z} S_{\alpha} (J_{\alpha}^{\rm exo} + J_{\alpha}^{\rm X} + J_{\alpha}^{\rm \star})\ , \label{eq:xa}
\end{equation}
where $S_{\alpha}$ is a quantum mechanical correction factor \cite{Furlanetto:2006jb}, $J_{\alpha}^{\rm X}$ and $J_{\alpha}^{\rm \star}$ stand for the fluxes of Lyman-$\alpha$ photons from astrophysical X rays and stellar emissions, respectively.
These quantities can be computed in detail using \texttt{21cmFAST} \cite{Mesinger:2010ne}. 
The contribution to the Lyman-$\alpha$ flux from exotic energy injection is given by  \cite{Mena:2019nhm} 
\begin{equation}
    J_{\alpha}^{\rm exo} = F_{\rm exc}(z) \frac{1}{n_{\rm b}} \frac{c n_{\rm b}}{4 \pi} \frac{1}{E_{\alpha}} \frac{1}{H(z) \nu_{\alpha}}\left(\frac{{\rm d} E}{{\rm d} V {\rm d} t}\right)_{\rm inj, L}\ , \label{eq:Jexo}
\end{equation}
where $F_{\rm exc}$ stands for the energy deposition efficiency through the process of hydrogen excitation, which can be calculated in detail using \texttt{DarkHistory} \cite{Liu:2019bbm}, $E_{\alpha}$ is the Lyman-$\alpha$ energy, and $\nu_{\alpha}$ is the Lyman-$\alpha$ frequency.

\section{Fisher-matrix forecasting}
\label{sec:Fisher}

To study the sensitivity of the Hongmeng project in probing \ac{DM} particles and \acp{PBH}, we employ the Fisher information matrix, which is particularly useful for theoretical analysis. 

\begin{figure}
    \centering
    \includegraphics[width=0.9\textwidth]{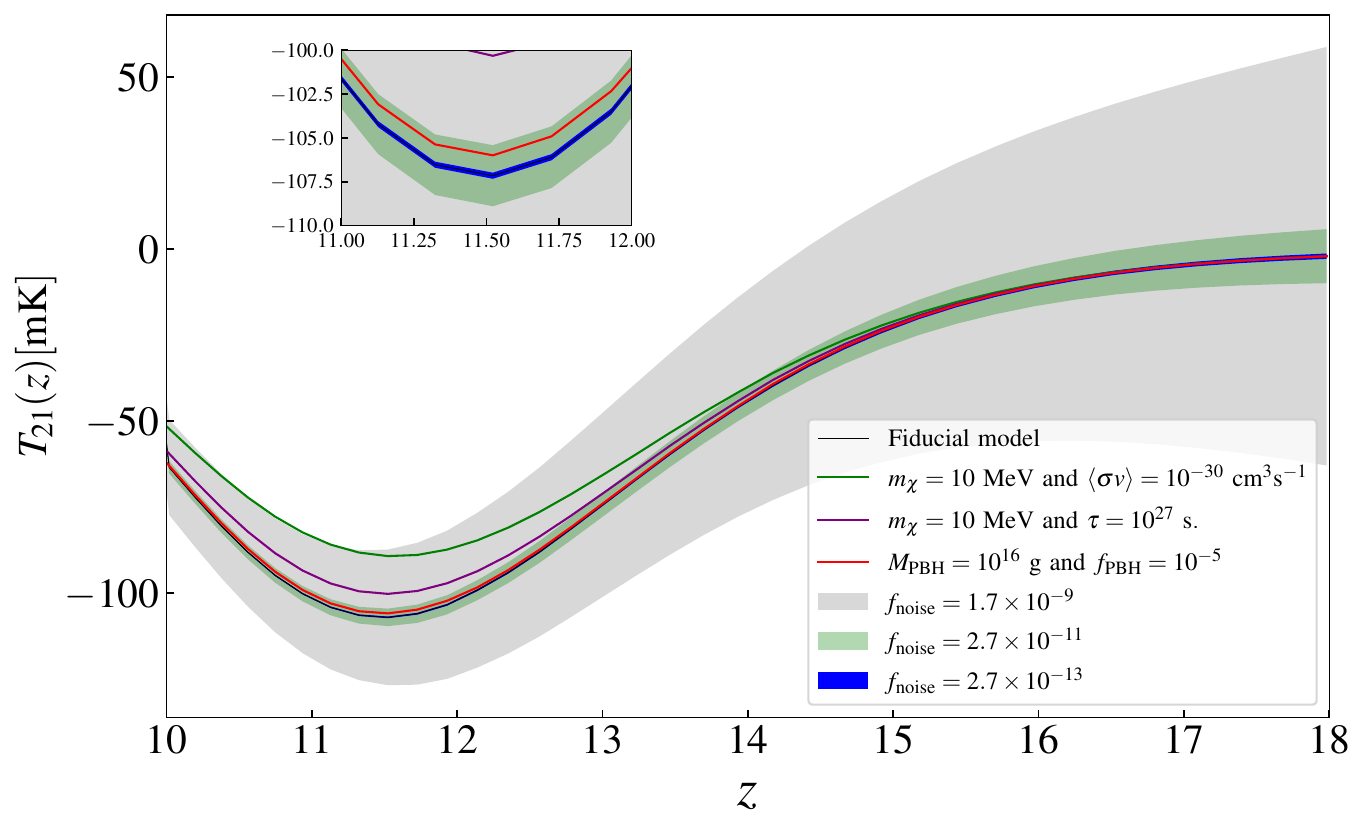}
    \caption{ {Prospective measurements of the Hongmeng mission for the 21 cm global spectrum.}
    {The gray, green and blue shaded regions show measurement errors with the noise factors of $f_{\rm noise} = 1.6 \times 10^{-9}$, $2.7 \times 10^{-11}$ and $2.7 \times 10^{-13}$, which correspond to the instrumental noise of integration time $t_{\rm int} = 600$\,seconds, $10$\,hours and $1000$\,hours, respectively, with an ideal foreground residual $\epsilon_{0} = 0$.}
    {The fiducial model of the 21 cm global signal is shown as the black curve.     
    Green curve shows 21 cm global signal with \ac{DM} annihilation through $\chi\chi\rightarrow e^{+}e^{-}$ channel with $ m_{\chi} = 10 ~\rm{MeV}$ and $\langle \sigma v \rangle = 10^{-29}~ \rm{cm}^{3}~s^{-1}$.
    Purple curve shows 21 cm global signal with \ac{DM} particle decay through $\chi\rightarrow e^{+}e^{-}$ channel with $ m_{\chi} = 10 ~\rm{MeV}$ and $\tau = 10^{27}~\rm{s}$.
    Red curve shows 21 cm global signal with \ac{PBH} contribution where $ M_{\rm PBH} = 10^{16} ~\rm{g}$ and $f_{\rm PBH} = 10^{-5}$.}
    }
    \label{fig:measurement error}
\end{figure}

We begin with the general form of the Fisher information matrix \cite{Tegmark:1996bz}
\begin{equation}
    F_{ij}=\frac{1}{2} {\rm Tr} \left[ C^{-1}C_{,i}C^{-1}C_{,j}+C^{-1}(\mu_{,i}\mu^T_{,j}+\mu_{,j}\mu^T_{,i}) \right] \ , \label{eq:Fisher}
\end{equation}
where $C$ stands for the covariance matrix of data, $C_{,i}$ denotes the derivative to the $i$-th parameter, denoted as $p_{i}$, and $\mu$ is the expectation value of an observable. 
The observable for 21 cm global experiments is the antenna temperature $T_{\rm sky}$ in a frequency band $\nu$, represented as \cite{Pritchard:2010pa,Liu:2019ygl,Facchinetti:2023slb}
\begin{equation}
    T_{\rm sky}(\nu) = T_{\rm fg}(\nu) +T_{\rm 21}(\nu)\ , \label{eq:Tsky}
\end{equation}
where $T_{\rm fg}$ and $T_{\rm 21}$ stand for the foreground temperature and differential 21 cm brightness temperature, respectively. 
In practice, the foreground temperature is characterized by a power-law, i.e., \cite{2023ChJSS..43...43C,Jester:2009dw}
\begin{equation}
    T_{\rm fg}(\nu) = 16.3 \times 10^6 \, {\rm K} \left(\frac{\nu}{2\, \rm MHz}\right)^{-2.53} \ .
    \label{eq:foregroundfit}
\end{equation}
The covariance is diagonal, since the detection in different frequency bands, labeled as $m$ and $n$, is expected to be uncorrelated. 
It takes the following form \cite{Pritchard:2010pa}
\begin{equation}
    C_{mn}=\delta_{mn}\sigma_{n}^{2} = \delta_{mn}  T_{\rm sky}^{2}(\nu_{n}) f_{\rm noise} \ . \label{eq:cover}
\end{equation}
Here, measurement error $\sigma_{n}$ consists of foreground residuals and thermal noise of instruments in the $n$-th frequency band. 
$f_{\rm noise}$ is a noise factor, which is 
\begin{equation}
    f_{\rm noise}  = 
    \frac{\epsilon^{2}_{0}\theta^{2}_{\rm fg}}{4\pi f_{\rm sky}} +\frac{1}{t_{\rm int} B} \ , \label{eq:sigma}
\end{equation}
where the first term on the right-hand side stands for the foreground residuals, and the second term corresponds to the thermal noise of instruments.
$\epsilon_{0}$ represents the fraction of foreground residuals in the signal, and $\theta_{\rm fg}$ is the angular resolution of the foreground model, which is set to $5^{\circ}$ \cite{Liu:2019ygl}.
The parameter $f_{\rm sky}$ denotes the sky-coverage fraction, which is fixed to $0.8$. 
In addition, $t_{\rm int}$ and $B$ represent the integration time and bandwidth, respectively. 
We consider several integration durations ranging from $600$ seconds to $1000$ hours, and the bandwidth is set to $1~\rm MHz$. 
Figure \ref{fig:measurement error} shows the measurement errors of the Hongmeng project for three noise factors.

Substituting \cref{eq:Tsky,eq:foregroundfit,eq:cover,eq:sigma} into \cref{eq:Fisher}, we obtain the specific form of Fisher information matrix for the Hongmeng project  \cite{Tegmark:1996bz,Pritchard:2010pa,Liu:2019ygl}
\begin{equation}
    F_{ ij}= \sum_{n=1}^{N_{\rm ch}}\left[2+f_{\rm noise}^{-1}\right]   \frac{{\rm d}\log T_{\rm sky}(\nu_{\rm n})}{{\rm d} p_{ i}} \frac{{\rm d}\log T_{\rm sky}(\nu_{\rm n})}{{\rm d} p_{ j}}\ . \label{eq:Fishermatrix}
\end{equation}
The signal is divided into $N_{\rm ch}$ frequency bands, each corresponding to a redshift bin. 
Details of these frequency bands are provided in Table~2 of Ref.~\cite{Facchinetti:2023slb}.

The model considered in this work is characterized by a set of independent parameters, i.e., 
\begin{equation}
\bold{p} = {\{ t_{\star}, a_{\star}, a_{\rm esc}, {\rm log}_{10}f_{\star}, {\rm log}_{10}f_{\rm esc},{\rm log}_{10}{L_{X}}, \langle \sigma {v} \rangle~{\rm or}~\Gamma~{\rm or}~f_{\rm PBH} \}}\ .
\end{equation} 
The first six parameters are astrophysical, following the conventions of \texttt{21cmFAST} \cite{Mesinger:2010ne}. 
In specific terms, the parameters in question are defined as follows: $t_{\star}$ represents a characteristic dimensionless star formation time scale, $a_{\star}$ is the exponent of a power-law stellar-to-halo mass ratio function, $a_{\rm esc}$ denotes the exponent of a power-law escape fraction function of stellar-emitted UV photons, ${\rm log}_{10}f_{\star}$ denotes the logarithmic coefficient of a power-law stellar-to-halo mass ratio function, ${\rm log}_{10}f_{\rm esc}$ stands for the logarithmic coefficient of the power-law escape fraction function for stellar-emitted UV photons, and ${\rm log}_{10}L_{\rm X}$ represents the logarithmic X-ray luminosity per unit star formation rate, in unit of $ \rm erg \cdot yr \cdot sec^{-1} M^{-1}_{\odot}$, where $M_{\odot}$ stands for the solar mass.
The remaining parameters correspond to \ac{DM} physics.
$\langle \sigma v \rangle$ characterizes the annihilation cross section of \ac{DM} particles, in unit of $ \rm cm^{-3}~s^{-1}$.
$\Gamma = \tau^{-1}$ represents the decay rate of \ac{DM} particles, in unit of $\rm s^{-1}$.
$f_{\rm PBH}$ signifies the abundance of \acp{PBH}.
In the fiducial model shown in figure \ref{fig:measurement error}, the astrophysical parameters are assumed to $t_{\star} = 0.5$, $a_{\star} = 0.5$, $a_{\rm esc} = -0.5$, ${\rm log}_{10}f_{\star} = -1.3$, ${\rm log}_{10}f_{\rm esc} = -1.0$, and ${\rm log}_{10}L_{\rm X} = 40.0$, while the other parameters, namely $f_{\rm PBH}$, $\langle \sigma v \rangle$, and $\Gamma$ are set to zero.

\section{Hongmeng's discovering potential}
\label{sec:RESULTS}

In this section, we present the prospective sensitivity of the Hongmeng project in probing \ac{DM}. 
We also compare its capability with that of other experiments. 

\subsection{Results for DM particles}

\begin{figure}
    \centering
    \includegraphics[width=1.0\textwidth]{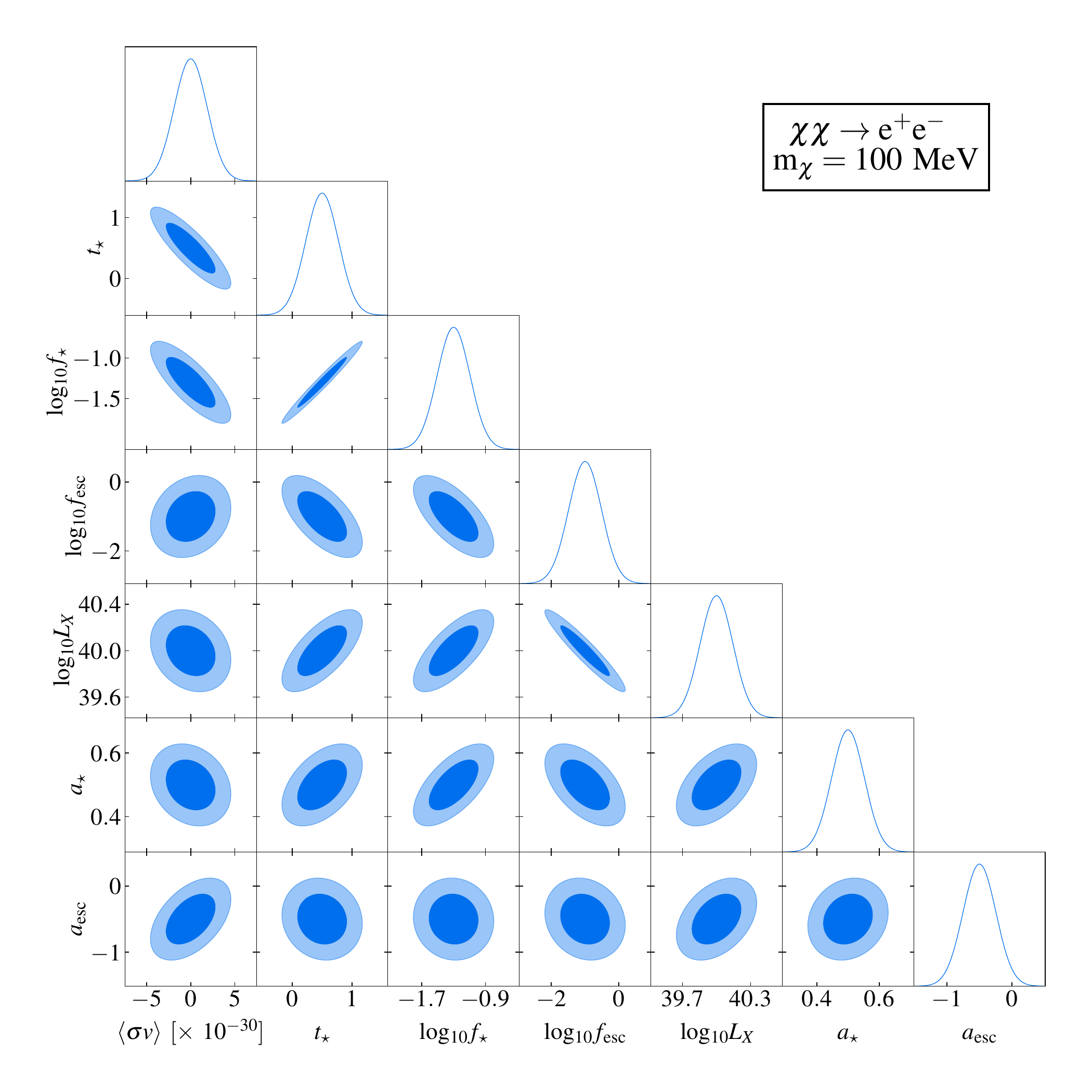}
    \caption{Fisher forecast for probing \ac{DM} annihilation through the $\chi \chi \rightarrow e^{+} e^{-}$ channel using the 21 cm global spectrum by the Hongmeng project.
    Dark and light shaded regions correspond to contours at $1\sigma$ and $2\sigma$ confidence intervals, respectively.
    Solid curves represent the marginalized posteriors of the model parameters.
    Fiducial model used is consistent with that shown in figure~\ref{fig:measurement error}.
    The mass of \ac{DM} particle is assumed to be $m_{\chi} = 100$\,MeV, with an integration duration of 1000 hours.
    }
    \label{fig:ann_corr}
\end{figure}

\begin{figure}
    \centering
    \includegraphics[width=0.5\textwidth]{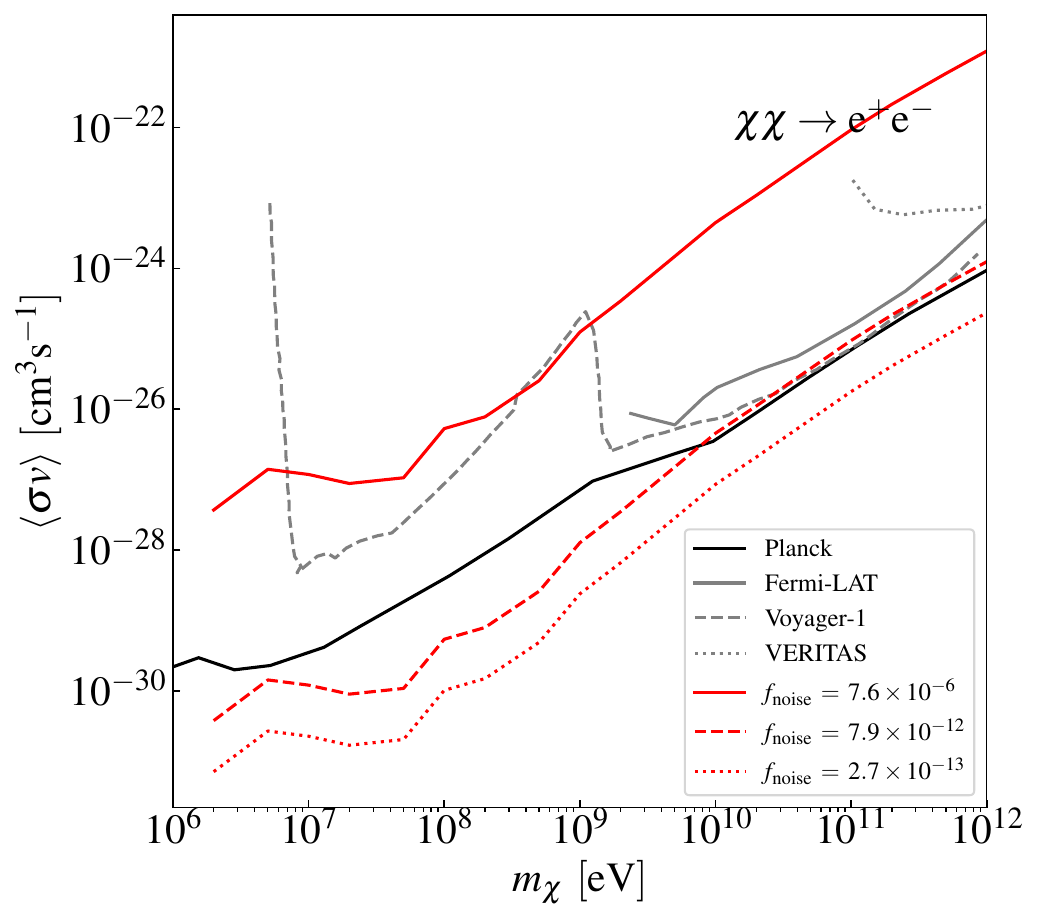}
    \includegraphics[width=0.5\textwidth]{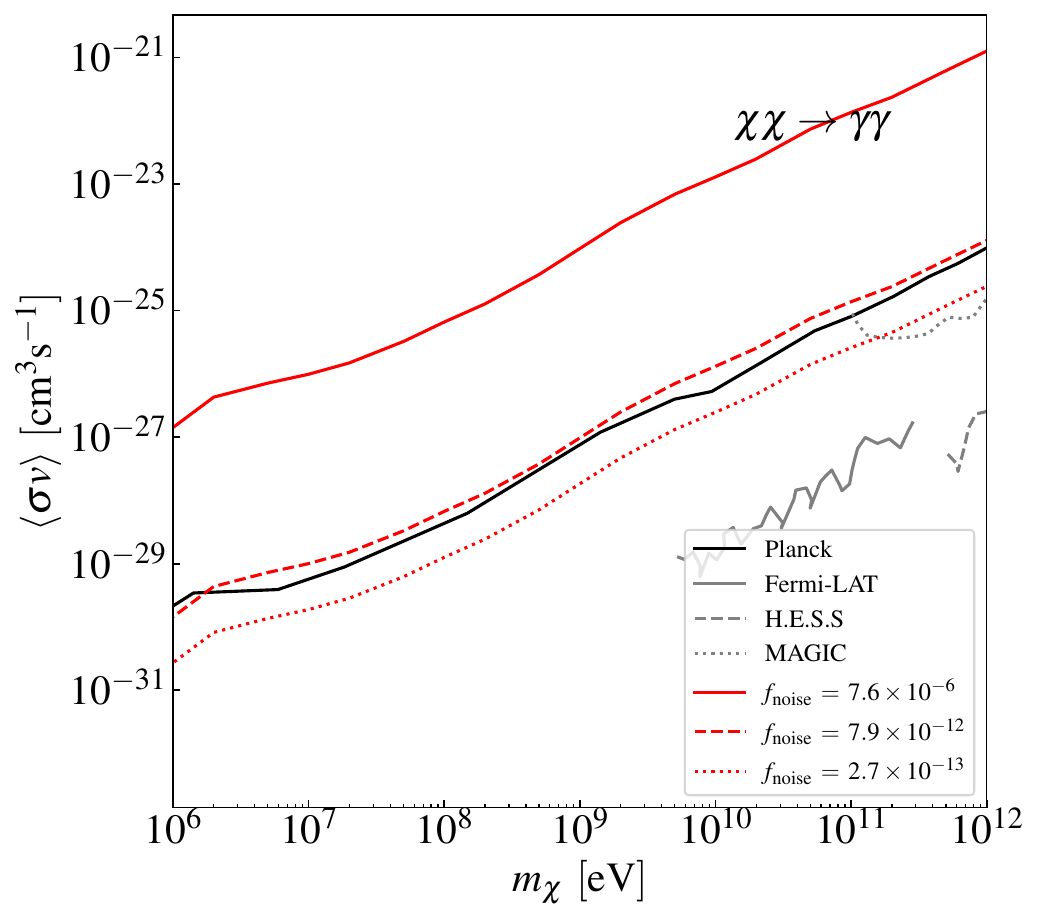}
    \includegraphics[width=0.5\textwidth]{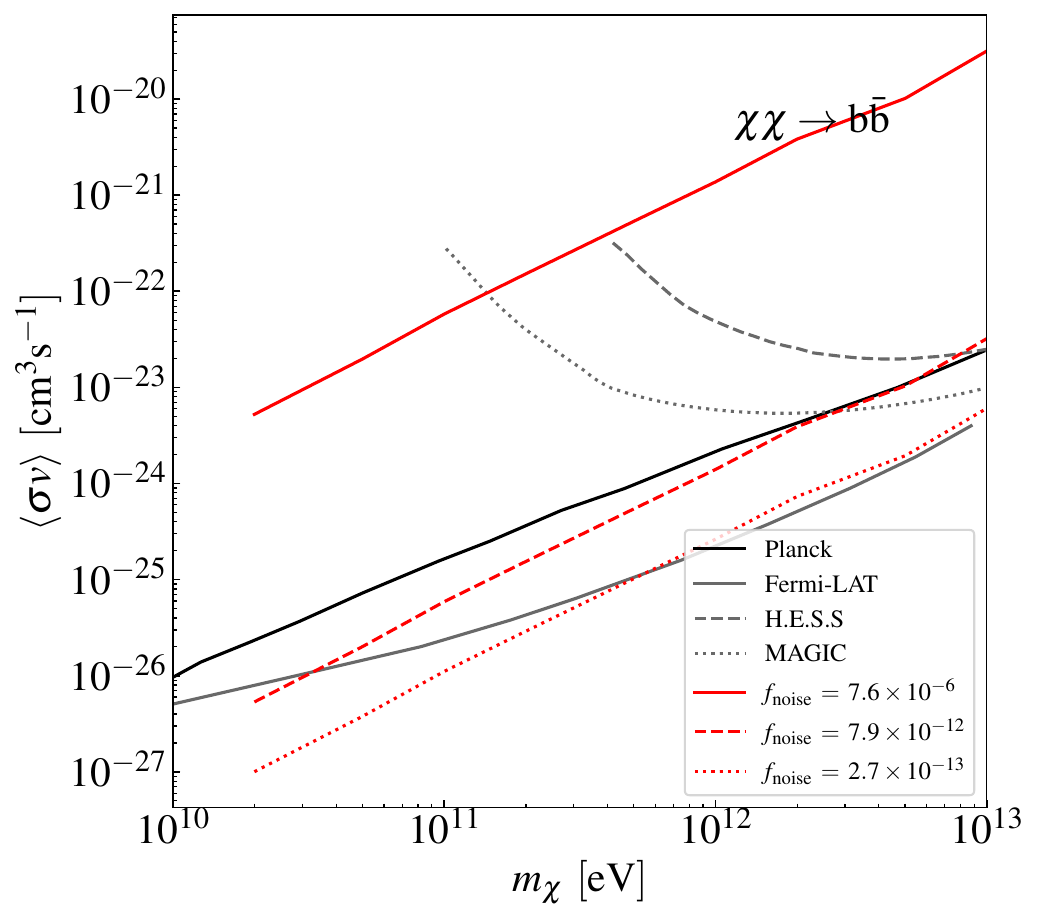}
    \caption{Prospective sensitivity of the Hongmeng project for probing the annihilation of \ac{DM} particles.
    The $1\sigma$ confidence-level sensitivity of the Hongmeng project to the thermally averaged annihilation cross-section of \ac{DM} particles (mass range $10^{6}$-$10^{12}\,\text{eV}$) is shown by the red curves.
    Existing $2\sigma$ upper limits from observations of \ac{CMB} distortion (black curve) \cite{Zhang:2023usm}, gamma-ray observations (gray curves) \cite{Cirelli:2020bpc,HESS:2018kom,HESS:2014zqa,VERITAS:2017tif,MAGIC:2017avy,Aleksic:2013xea,Fermi-LAT:2015att}, and electron-positron pairs (gray dashed curve) \cite{Cohen:2016uyg,Boudaud:2018hqb} are included for comparison.
    }
    \label{fig:ann_sense}
\end{figure}

\begin{figure}
    \centering
    \includegraphics[width=1.0\textwidth]{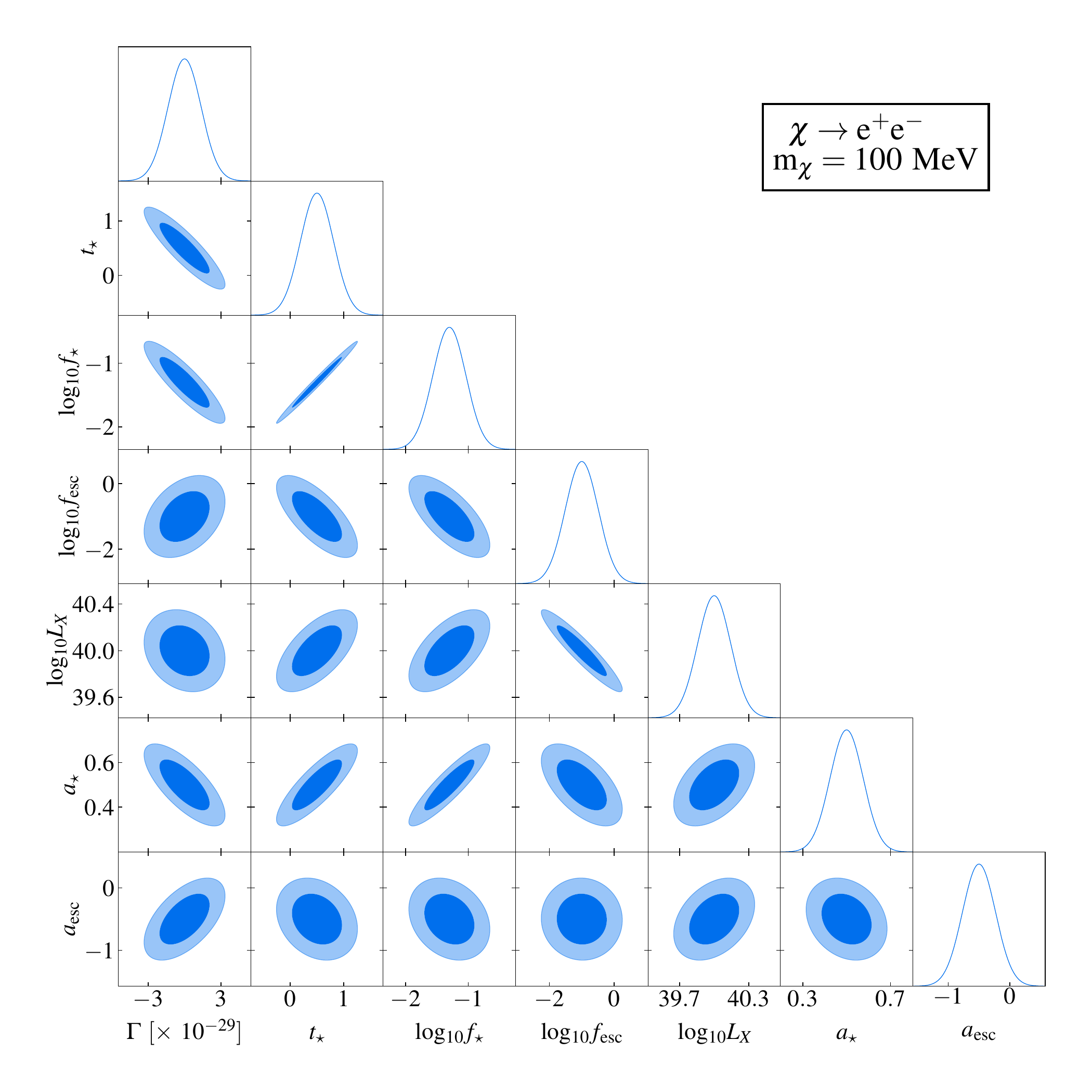}
    
    \caption{Fisher forecast for probing \ac{DM} decay through the $\chi \rightarrow e^{+} e^{-}$ channel using the 21 cm global spectrum by the Hongmeng project.
    Dark and light shaded regions correspond to contours at $1\sigma$ and $2\sigma$ confidence intervals, respectively.
    Solid curves represent the marginalized posteriors of the model parameters.
    Fiducial model used is consistent with that shown in figure~\ref{fig:measurement error}.
    The mass of \ac{DM} particle is assumed to be $m_{\chi} = 100$\,MeV, with an integration duration of 1000 hours.}
    \label{fig:dec_corr}
\end{figure}

\begin{figure}
    \centering
    \includegraphics[width=0.5\textwidth]{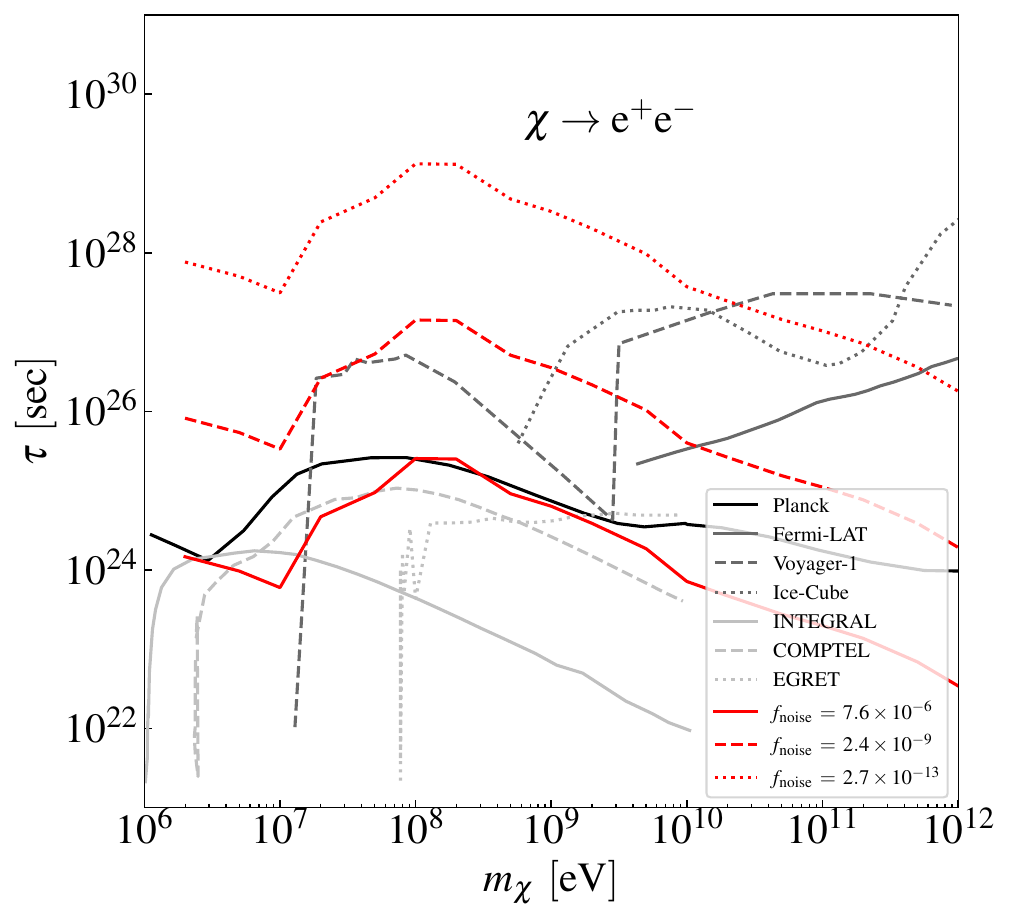}
    \includegraphics[width=0.5\textwidth]{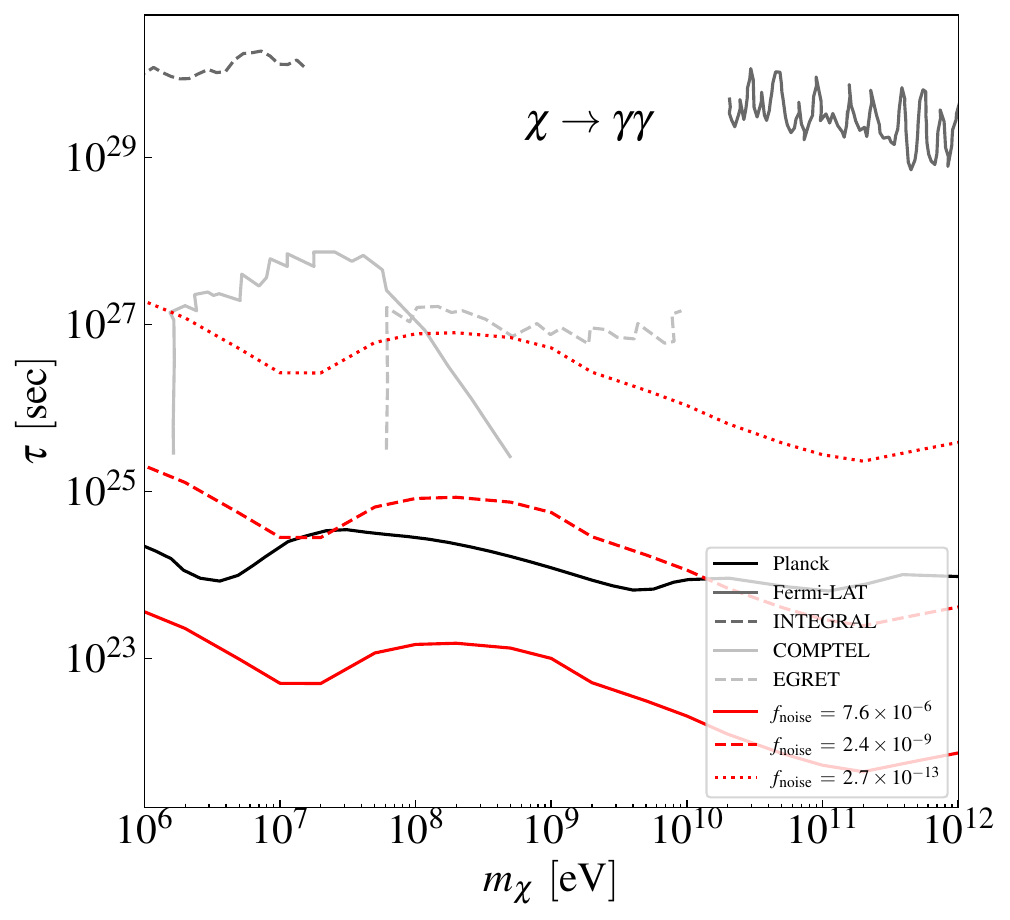}
    \includegraphics[width=0.5\textwidth]{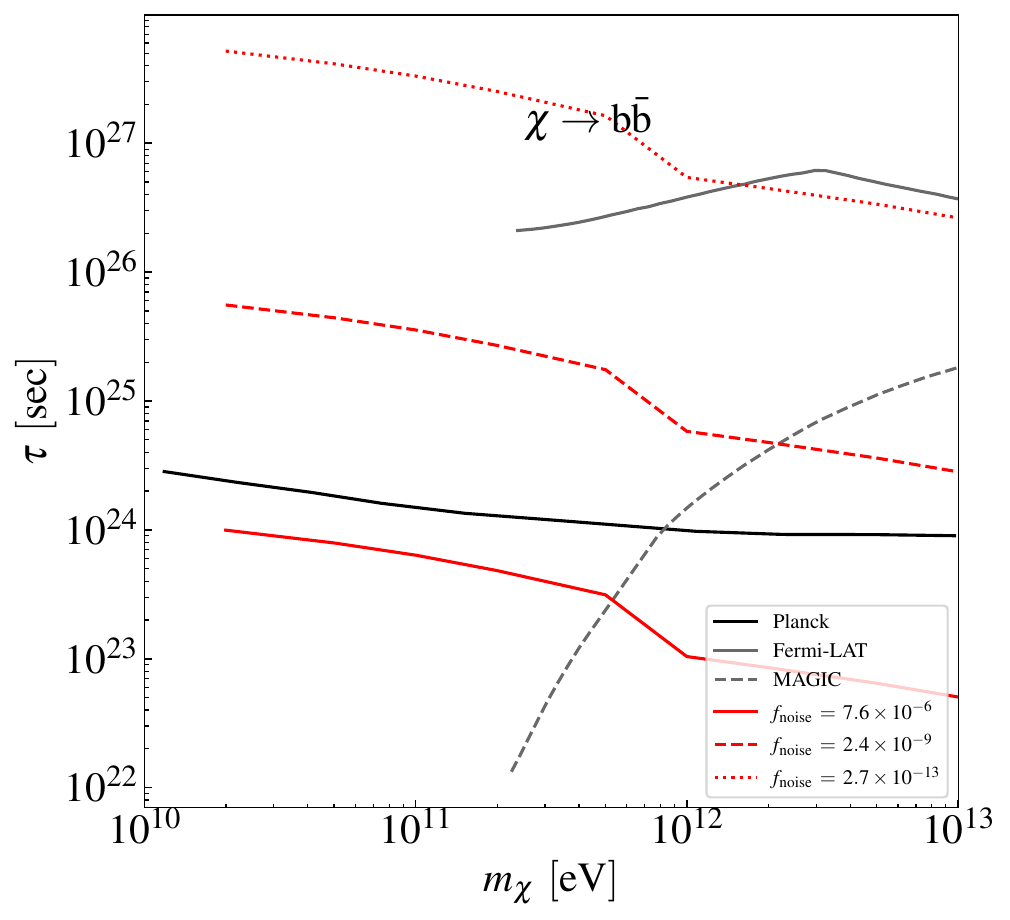}
    \caption{Prospective sensitivity of the Hongmeng project for probing the decay of \ac{DM} particles.
    The $1\sigma$ confidence-level sensitivity of the Hongmeng project to the decay lifetime of \ac{DM} particles (mass range $10^{6}$-$10^{12}\,\text{eV}$) is shown by the red curves.
    Existing $2\sigma$ upper limits from observations of \ac{CMB} distortion (black curve) \cite{Planck:2018vyg,Slatyer:2015kla,Capozzi:2023xie}, extragalactic photons (gray curves) \cite{Essig:2013goa,Cohen:2016uyg,Massari:2015xea,Cadamuro:2011fd,Koechler:2023ual,Calore:2022pks,Cirelli:2020bpc,Foster:2022nva}, and electron-positron pairs (gray dashed curve) \cite{Cohen:2016uyg,Boudaud:2018hqb,Cohen:2016uyg} are included for comparison.}
    \label{fig:dec_sense}
\end{figure}

{We present the Fisher matrix analysis results in figures~\ref{fig:ann_corr}--\ref{fig:dec_sense}.}
{\Cref{fig:ann_corr,fig:dec_corr} present two representative results of Fisher matrix analysis, which illustrate the correlations among and constraints on model parameters, assuming a \ac{DM} particle mass of $m_{\chi}=100$\,MeV and an integration duration of $1000$\,hours.
The complete triangle plots are provided in \Cref{sec:appendix}.}
The dark and light shaded regions represent two-dimensional contours at $1\sigma$ and $2\sigma$ confidence levels, respectively, while the solid curves represent one-dimensional marginalized posterior probability distribution functions.
Here, we adopt the same fiducial model as in figure~\ref{fig:measurement error}. 
{
\Cref{fig:ann_sense,fig:dec_sense} show the prospective $1\sigma$-confidence-level sensitivity of the Hongmeng project to probe the annihilation or decay of \ac{DM} particles across the mass range of $10^{6}-10^{12}$\,eV, considering various noise factors.
In this work, for the conservative scenario, we adopt $\epsilon_{0} = 0.1$ and $t_{\rm int} = 600$\,seconds, corresponding to a noise factor $f_{\rm noise} = 7.6 \times 10^{-6}$ and a measurement error $\sigma_{n}$ about $1000$\,mK at $120$\,MHz.
According to Ref.~\cite{deOliveira-Costa:2008cxd}, $\epsilon_{0} = 0.1$ has already been achieved,  making it a reasonable conservative estimate of the foreground residual.
An optimistic estimate assumes $\epsilon_{0} = 0$ and $t_{\rm int} = 1000$\,hours, yielding a noise factor $f_{\rm noise} = 2.7 \times 10^{-13}$ and a corresponding measurement error $\sigma_{n}$ about $0.2$\,mK at $120$\,MHz.}
For comparison, we show existing $2\sigma$ upper limits from observations of the \ac{CMB} distortions (Planck 2018, black curves) \cite{Zhang:2023usm}, gamma rays (\ac{Fermi-LAT}, \ac{H.E.S.S},\ac{VERITAS}, \ac{MAGIC}, gray curves) \cite{Fermi-LAT:2015att,Cirelli:2020bpc,HESS:2018kom,HESS:2014zqa,VERITAS:2017tif,MAGIC:2017avy,Aleksic:2013xea}, and electron-positron pairs (Voyager-1, gray dashed curve) \cite{Boudaud:2016mos,Boudaud:2018oya}.

{Figures \ref{fig:ann_corr} and \ref{fig:dec_corr} reveal subtle correlations between the annihilation or decay parameters of \ac{DM} particles and the astrophysical parameters.
This indicates a limited degeneracy between the exotic energy injection from the \ac{DM} annihilation and decay, and astrophysical processes affecting the 21 cm global signal.
{Specifically, $\langle \sigma v \rangle$ and $\tau$ are positively correlated with $a_{\rm esc}$ and ${\rm log}_{10}f_{\rm esc}$ , and negatively correlated with $t_{\star}$ and ${\rm log}_{10}f_{\star}$.}
Consequently, it is feasible to simultaneously constrain both the \ac{DM} and astrophysical parameters.
This, in turn, enables the extraction of key \ac{DM} properties, such as their thermally-averaged annihilation cross section $\langle \sigma v \rangle$ and lifetime $\tau$.}

{Figure \ref{fig:ann_sense} reveals that the Hongmeng project demonstrates enhanced sensitivity in probing \ac{DM} particles through the annihilation channel $\chi\chi\rightarrow e^{+}e^{-}$ compared to other channels like $\chi\chi\rightarrow\gamma\gamma$ and $\chi\chi\rightarrow b \bar{b}$.  
Focusing on the optimal annihilation channel $\chi\chi\rightarrow e^{+}e^{-}$, as shown in the top panel of figure \ref{fig:ann_sense},
we find that with a noise factor of $f_{\rm noise} \sim 10^{-11}$, as indicated by the red dashed curve, the Hongmeng project's sensitivity is comparable to the most stringent constraints, shown as gray solid curves.
This suggests that the Hongmeng project is poised to test existing results in the near future.
The red curve comparison indicates that the sensitivity can be improved by reducing the noise factor.
Referring to \cref{eq:sigma}, the noise factor can be reduced by minimizing foreground residuals or increasing the integration time.
From \cref{eq:sigma}, we find that for the red dashed curve that for the red dashed curve, a foreground residual of $\epsilon_{0}=10^{-4}$ and thermal noise over $16.7$\,hours contribute equally to the total measurement error of the Hongmeng project.
When $\epsilon_{0} \gtrsim 10^{-4}$ (e.g., $\epsilon_{0}=10^{-3}$), foreground residuals dominate the measurement error.
In this regime, reducing the residual is key to improving sensitivity, as it becomes the primary limiting factor over thermal noise.
Thus, efficient foreground subtraction methods are essential for future observations and data analysis.
Extending $t_{\rm int}$ also improves the sensitivity.
When $t_{\rm int} \lesssim 16.7$\,hours, (e.g., $t_{\rm int} = 600$\,seconds), thermal noise dominates, beyond this, foregrounds become the limiting factor.
As \cref{eq:sigma} indicates, in this case, the noise scales as $\sigma_{n}\propto t_{\rm int}^{-1/2}$.
Hence, longer integration times significantly improve the project's sensitivity. 
Additionally, the Hongmeng project is also expected to probe lighter \ac{DM} particles, particularly in the sub-GeV mass regimes, beyond the reach of current experiments.}

Based on figure \ref{fig:dec_sense}, we find that the Hongmeng project is more sensitive in probing \ac{DM} particles via the decay channel $\chi\rightarrow e^{+}e^{-}$ than in probing $\chi\rightarrow\gamma\gamma$ and $\chi\rightarrow b \bar{b}$.  
To clarify this result, we focus on the optimal decay channel $\chi\rightarrow e^{+}e^{-}$, as shown in the top panel of figure \ref{fig:dec_sense}. 
{
We present the sensitivity of the Hongmeng project under different noise factors.
The red solid, dashed, and dotted curves represent the sensitivity for noise factor of $f_{\rm noise} \sim 10^{-5}$, $10^{-9}$, and $10^{-13}$ respectively.
The red curves indicate that reducing the noise factor improves the sensitivity.}
{As shown in \cref{eq:sigma}, the noise factor can be reduced by minimizing foreground residuals or increasing the integration time.
For example, with $\epsilon_{0}=10^{-3}$ and  $t_{\rm int} = 600$\,seconds, foreground residual and thermal noise contribute equally to the total measurement error.
When $\epsilon_{0} \gtrsim 10^{-3}$ (e.g., $\epsilon_{0}=10^{-2}$), foreground residuals dominate the measurement error.
In this regime, reducing foreground contamination becomes essential for improving sensitivity.
Extending $t_{\rm int}$ also improves the sensitivity.
When $t_{\rm int} \lesssim 600$\,seconds, the instrumental noise would dominate the measurement error.
In this case, we have $\sigma_{n}\propto t_{\rm int}^{-1/2}$.
Therefore, extending the integration duration would effectively improve the sensitivity of the Hongmeng project.
}

{We also compare the prospective sensitivity of the Hongmeng project with the existing upper limits on $\tau$. 
At $f_{\rm noise} \sim 10^{-9}$, the sensitivity of the Hongmeng project is comparable to the tightest current constraints (black and gray curves), suggesting that existing bounds could be tested in the near future.
When considering lower noise factor, we show that the sensitivity of the Hongmeng project can be further enhanced, as mentioned above. 
Furthermore, compared to existing experiments, the Hongmeng project may probe the decay of lighter \ac{DM} particles, particularly in the sub-GeV mass regimes.}

\subsection{Results for PBHs}

\begin{figure}
    \centering
    \includegraphics[width=1.0\linewidth]{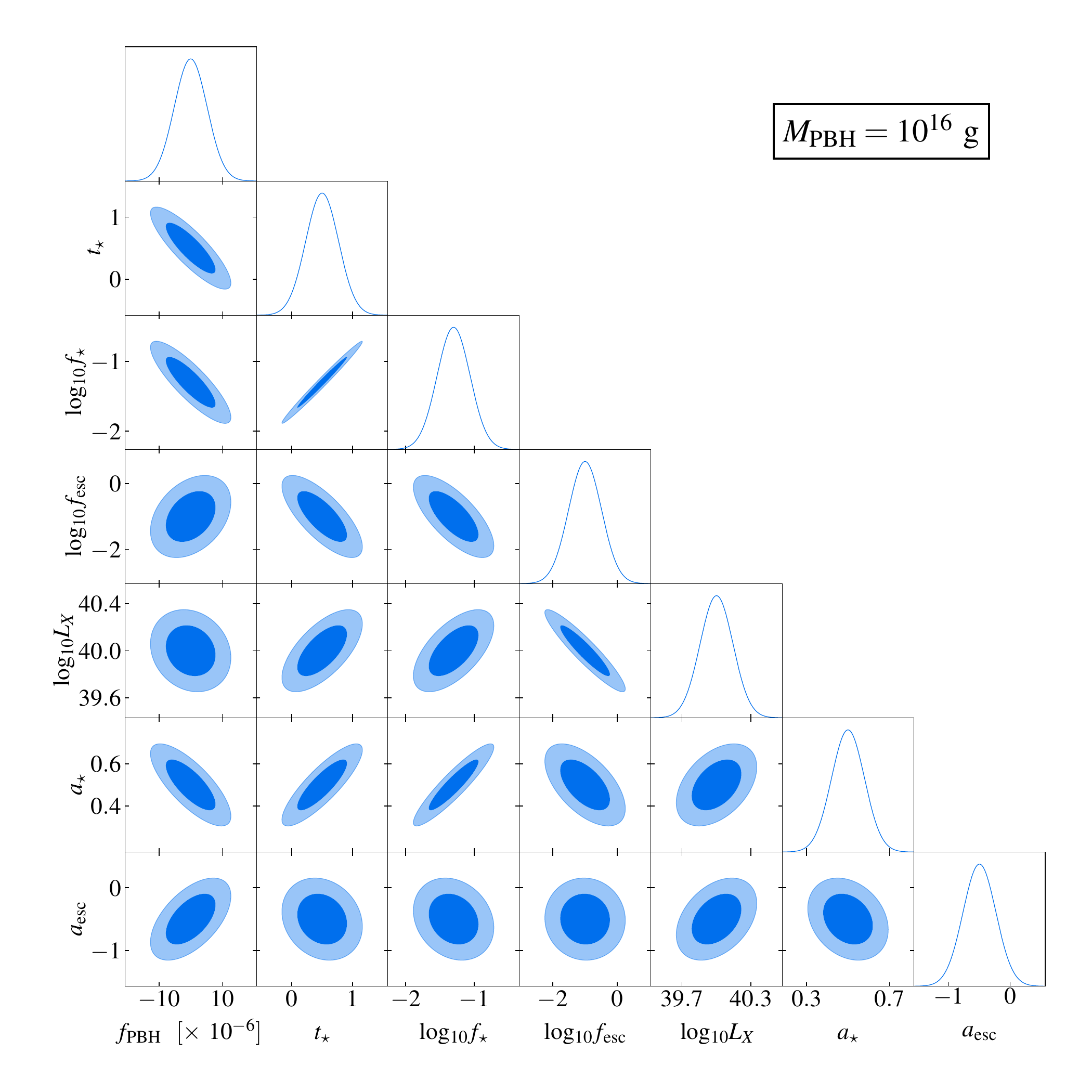}
    \caption{Fisher forecast for probing \ac{PBH} Hawking radiation using the 21 cm global spectrum by the Hongmeng project.
    Dark and light shaded regions correspond to contours at $1\sigma$ and $2\sigma$ confidence intervals, respectively.
    Solid curves represent the marginalized posteriors of the model parameters.
    Fiducial model used is consistent with that shown in figure~\ref{fig:measurement error}.
    The mass of \ac{PBH} is assumed to be $M_{\rm PBH} = 10^{16}$\,g, with an integration duration of 1000 hours.}
    \label{fig:PBH_corr}
\end{figure}

\begin{figure}
    \centering
    \includegraphics[width=0.8\linewidth]{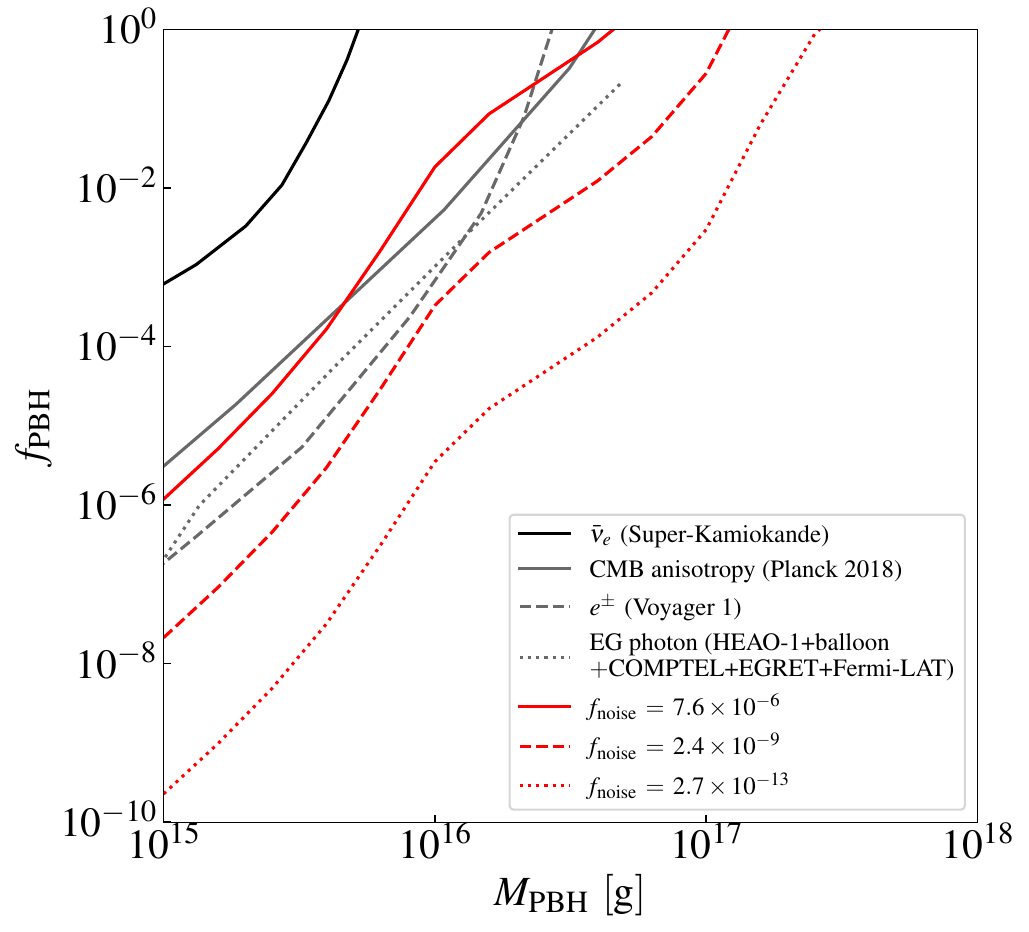}
    \caption{Prospective sensitivity of the Hongmeng project for probing the \acp{PBH}.
    $1\sigma$-confidence-level sensitivity of the Hongmeng project to measure the abundance of \acp{PBH} within the mass range of $10^{15}-10^{18}$\,g are shown by the red curves.
    For comparison, we show the existing upper limits at $2\sigma$ confidence level from observations of the diffusion neutrino background (black curve) \cite{Poulin:2016anj,Wang:2020uvi}, \ac{CMB} anisotropies (gray solid curve) \cite{,Chluba:2020oip,Acharya:2020jbv,Clark:2016nst}, extra-galactic photons (gray dotted curve) \cite{Carr:2009jm,Carr:2016hva}, and electron-positron pairs (gray dashed curve) \cite{Cohen:2016uyg,Boudaud:2018hqb}. }
    \label{fig:PBH_sense}
\end{figure}

We summarize the results of Fisher-matrix analysis in figures \ref{fig:PBH_corr} and \ref{fig:PBH_sense}. 
In figure \ref{fig:PBH_corr}, we present triangle plots showing the correlations and constraints on model parameters, considering \acp{PBH} with masses of $10^{16}$\,g and an integration duration of 1000 hours. 
Specifically, the dark and light shaded regions indicate the $1\sigma$ and $2\sigma$ two-dimensional confidence contours, while the solid curves represent one-dimensional marginalized posterior probability distributions. 
Here, we adopt the same fiducial model as in figure \ref{fig:measurement error}. 
In figure \ref{fig:PBH_sense},  we present the projected $1\sigma$ sensitivity of the Hongmeng project in searching for \ac{PBH} Hawking radiation within the mass range $10^{15}-10^{18}$\,g, under different assumptions of noise factors. 
For comparison, we further depict the existing upper limits at $2\sigma$ confidence level from observations of the diffusion neutrino background (Super-Kamionkande, black curve) \cite{Poulin:2016anj,Wang:2020uvi}, \ac{CMB} anisotropies (Planck 2018 results, gray solid curve) \cite{Carr:2016hva,Chluba:2020oip,Acharya:2020jbv,Clark:2016nst}, extra-galactic photons (a combination of the \ac{HEAO}, the \ac{COMPTEL}, the \ac{EGRET}, and the \ac{Fermi-LAT}, gray curves) \cite{Carr:2009jm,Carr:2016hva}, and electron-positron pairs (Voyager-1, gray dashed curve) \cite{Cohen:2016uyg,Boudaud:2018hqb}. 

\iffalse
\begin{figure}
\centering
\includegraphics[width=0.47\textwidth]{pbhcorrlation.pdf}
\quad
\includegraphics[width=0.47\textwidth]{f_pbh_new.pdf}
\caption{Prospective sensitivity of the Hongmeng project for searching the Hawking radiation of \acp{PBH}. Left panel: Triangle plots of the model parameters. The dark and light shaded regions, respectively, represent contours at $1\sigma$ and $2\sigma$ confidence levels, while the solid curves stand for marginalized posteriors of model parameters. The fiducial model is the same as that of figure \ref{fig:measurement error}. For illustration, we consider the mass of \acp{PBH} of $M_{\rm PBH}=10^{16}$\,g. We set the integration duration to be 1000 hours. Right panel: The $1\sigma$-confidence-level sensitivity of the Hongmeng project (purple, red, and blue curves) to measure the abundance of \acp{PBH} in the mass range of $10^{15}-10^{18}$\,g. For comparison, we show the existing upper limits at $2\sigma$ confidence level from observations of the diffusion neutrino background (Super-Kamionkande, black curve) \cite{Poulin:2016anj,Wang:2020uvi}, \ac{CMB} anisotropies (Planck 2018 results, gray solid curve) \cite{,Chluba:2020oip,Acharya:2020jbv,Clark:2016nst}, extra-galactic photons (\ac{HEAO}, \ac{COMPTEL}, \ac{EGRET}, and Fermi-LAT, gray dashed curve) \cite{Carr:2009jm,Carr:2016hva}, and electron-positron pairs (Voyager-1, green curve) \cite{Cohen:2016uyg,Boudaud:2018hqb}. }
\label{fig:BHcor}
\end{figure}
\fi

{Figure \ref{fig:PBH_corr} reveals weak correlations between the parameter $f_{\rm PBH}$ and the astrophysical parameters. 
This suggests a low level of degeneracy between the exotic energy injection from \ac{PBH} Hawking radiation and the astrophysical processes that shape the 21 cm global signal.
Therefore, it is possible to simultaneously constrain both $f_{\rm PBH}$ and the astrophysical parameters. 
In principle, this enables the extraction of information on \acp{PBH}, such as their mass function.}

{Figure \ref{fig:PBH_sense} is used to analyze how the noise factor affects the projected sensitivity of the Hongmeng project in probing \acp{PBH}.  
The red curves indicate that reducing the noise factor improves sensitivity.  
According \cref{eq:sigma}, the noise factor can be reduced by minimizing foreground residuals or increasing the integration time.  
For example, the red dashed curve shows that when $\epsilon_{0}=10^{-3}$ and $t_{\rm int} = 600$\,seconds, foreground residuals and thermal noise contribute equally tothe total error. 
When $\epsilon_{0} \gtrsim 10^{-3}$ (e.g., $\epsilon_{0}=10^{-2}$), foreground residuals dominate the measurement error.
Thus, reducing them becomes essential for improving the project's sensitivity.
Efficient foreground subtraction techniques are therefore crucial for future observations and data analysis.
Extending $t_{\rm int}$ also improves sensitivity.  
For $t_{\rm int} \lesssim 600$\,seconds, instrumental noise dominates the measurement error, in this case we have $\sigma_{n}\propto t_{\rm int}^{-1/2}$.
Thus, extending the integration duration is essential for enhancing sensitivity.  
In summary, both reducing foreground residuals and ensuring sufficient integration time are critical for optimizing the sensitivity of the Hongmeng project.}

{
We compare our results with existing astronomical upper limits, as shown in figure \ref{fig:PBH_sense}. 
With a noise factor of $f_{\rm noise} \sim 10^{-9}$, the Hongmeng project achieves sensitivity comparable to the most stringent current constraints,represented by the gray curves.
This suggests that existing limits could be tested in the near future. 
Lower foreground residuals and longer integration times can further enhance the project's sensitivity, as demonstrated earlier. 
For instance, the Hongmeng project can probe down to $f_{\rm PBH}\simeq10^{-9}$ for \acp{PBH} of mass $10^{15}$\,g with a noise factor $f_{\rm noise} \sim 10^{-13}$. 
Additionally, compared to existing experiments, the Hongmeng project is expected to probe heavier \acp{PBH}, particularly in the $10^{17}$\,g range, which is otherwise inaccessible.  
}

\section{Summary}
\label{sec:CONCLUSION}

This study has explored the potential sensitivity of the Hongmeng project in probing the annihilation, decay of \ac{DM} particles, and the Hawking radiation from \acp{PBH} through observations of the 21 cm global spectrum.
It is anticipated that these processes would introduce exotic energy into the \ac{IGM}, leading to changes in the 21 cm global signal during cosmic dawn.
By employing Fisher-matrix analysis, we have evaluated the expected sensitivity of the Hongmeng project to the relevant model parameters, offering valuable insights for future experimental planning and data analysis.
Our findings indicate that the Hongmeng project is well-positioned to probe \ac{DM} particles and \acp{PBH} in the near future.

The Hongmeng project has the potential to probe the annihilation of \ac{DM} particles, specifically focusing on the channel that produces positron-electron pairs, which offers the highest sensitivity for probing.
However, in order for the Hongmeng project to reach sensitivity levels comparable to current astronomical observational constraints, it would need to address challenges such as foreground residuals of around $10^{-4}$ and integration times of approximately $10^{3}$ hours.
Failure to meet these requirements could impede the project's ability to achieve its objectives. Therefore, significant technical challenges must be overcome in the future to ensure successful detection in this regard.

The Hongmeng project is capable of observing the decay of \ac{DM} particles, particularly focusing on the decay channel that results in positron-electron pairs, offering superior sensitivity.
To achieve sensitivity levels comparable to current astronomical observational limits, the project would need to address challenges such as foreground residuals of approximately $10^{-3}$ and an integration duration of $600$ seconds.
Enhancing sensitivity practically could be achieved by extending the integration duration. Therefore, through the examination of the 21 cm global signal, the Hongmeng project shows potential for probing the decay of \ac{DM} particles within the sub-GeV mass range, surpassing current observational limits in sensitivity.
However, the implementation of effective foreground subtraction methods remains crucial for the success of these efforts.

In contrast, probing Hawking radiation from \acp{PBH} is relatively more straightforward for the Hongmeng project.
To achieve sensitivity levels comparable to current astronomical observational limits, the project would need to address challenges such as foreground residuals of around $10^{-3}$ and an integration duration of 600 seconds.
Practical enhancements in sensitivity could be achieved by extending the integration duration.
In comparison to existing astronomical experiments, the Hongmeng project has the ability to probe more massive \acp{PBH}, a feat that proves challenging for other experiments.

{For simplicity, we have assumed that model parameter biases are negligible. However, it is worth noting that foreground residuals generally introduce non-negligible biases \cite{Pattison:2023eej,Liu:2019awk,Deshpande:2018ugh}, which may affect parameter estimation.
The influence of such biases will be addressed in future work.
According to the error estimation framework proposed by de Oliveira-Costa \cite{deOliveira-Costa:2008cxd}, a conservative estimate for the foreground residual fraction is $\epsilon_{0}=0.1$.
Our analysis shows that achieving sensitivity comparable to existing constraints requires reducing $\epsilon_{0}$ to below $10^{-3}$.  
However under realistic observational conditions, current foreground subtraction techniques typically achieve  $\epsilon_{0} \sim 0.01$ \cite{Pattison:2023eej,Liu:2019awk,Deshpande:2018ugh}.
These findings highlight the urgent need for advanced subtraction techniques, including artificial intelligence-based approaches \cite{Tripathi:2024iai}.
The practical implementation of such methods will be explored in future work.
}

\iffalse
{It is worth noting that foreground residuals introduce not only significant statistical errors but also unwanted biases in model parameters.  
In this work, we assume these biases to be negligible, as they are substantially smaller than the statistical errors \cite{Pattison:2023eej,Liu:2019awk,Deshpande:2018ugh} and are not the primary focus of our investigation.  
A detailed examination of biases originating from foreground residuals will be addressed in future works.  
Furthermore, as demonstrated by the error estimation framework developed by de Oliveira-Costa in 2008, a conservative foreground residual fraction is $\epsilon_{0}=0.1$, which corresponds to a noise factor of $f_{\rm noise} \sim 10^{-5}$, as indicated by the red solid lines in our work.  
Idealized models suggest the theoretical possibility of $\epsilon_{0}=0$.  
Our results show that achieving a sensitivity comparable to existing constraints requires a foreground residual of less than $10^{-3}$.  
However, under realistic instrumental conditions, current state-of-the-art subtraction techniques achieve at best $\epsilon_{0}=0.01$.  
This underscores the urgent need for developing more efficient foreground subtraction methods.}
Given the stringent requirements for foreground subtraction in \ac{DM} probing, it is crucial to emphasize the importance and urgent need for the development of effective foreground subtraction methods, such as artificial intelligence \cite{Tripathi:2024iai}.
However, the specifics of this research will be left for future consideration.
\fi

\appendix
\section{Full results of Fisher matrix analysis}

\label{sec:appendix}

\begin{figure}
    \includegraphics[width=1.0\textwidth]{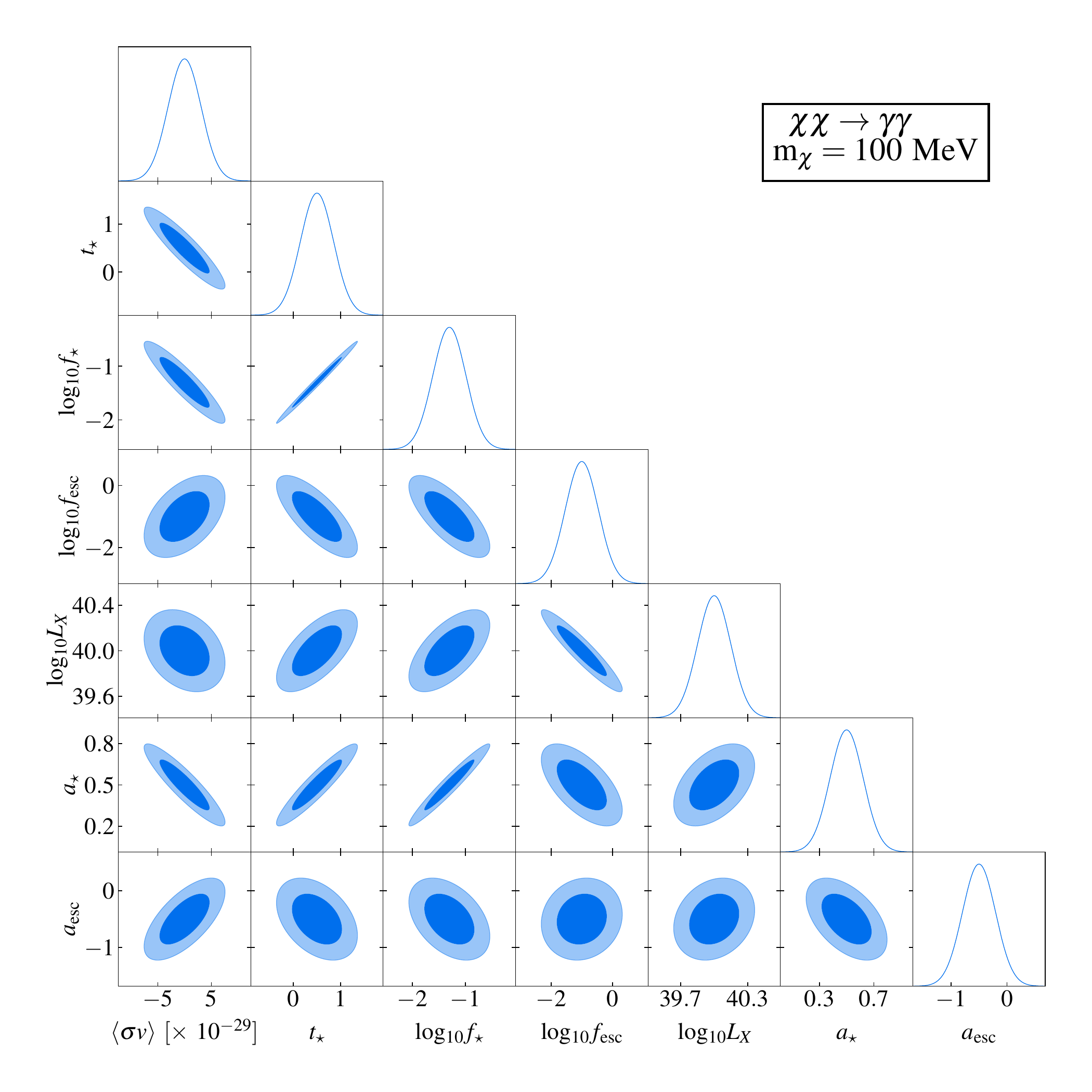}
    \caption{Fisher forecast for probing \ac{DM} annihilation through the $\chi \chi \rightarrow \gamma \gamma$ channel using the 21 cm global spectrum by the Hongmeng project.
    Dark and light shaded regions correspond to contours at $1\sigma$ and $2\sigma$ confidence intervals, respectively.
    Solid curves represent the marginalized posteriors of the model parameters.
    Fiducial model used is consistent with that shown in figure~\ref{fig:measurement error}.
    The mass of \ac{DM} particle is assumed to be $m_{\chi} = 100$\,MeV, with an integration duration of 1000 hours.}
\label{fig:ann_corry}
\end{figure}
    
\begin{figure}
    \includegraphics[width=1.0\textwidth]{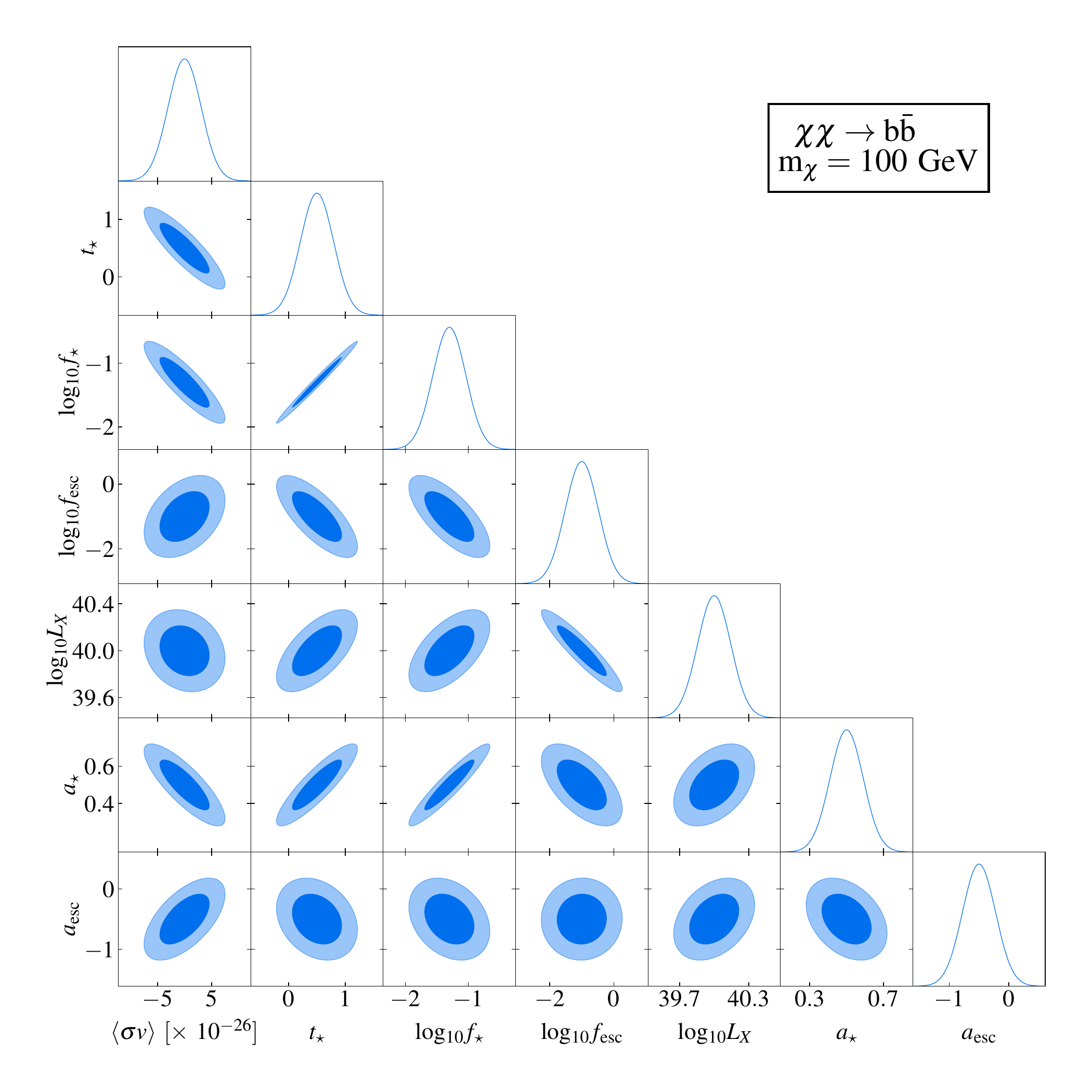}
    \caption{Same as figure \ref{fig:ann_corry} but for $100$\,GeV \ac{DM} particles annihilating into bottom-anti-bottom quark pairs.
    }
    \label{fig:ann_corrb}
\end{figure}

\begin{figure}
\includegraphics[width=1.0\textwidth]{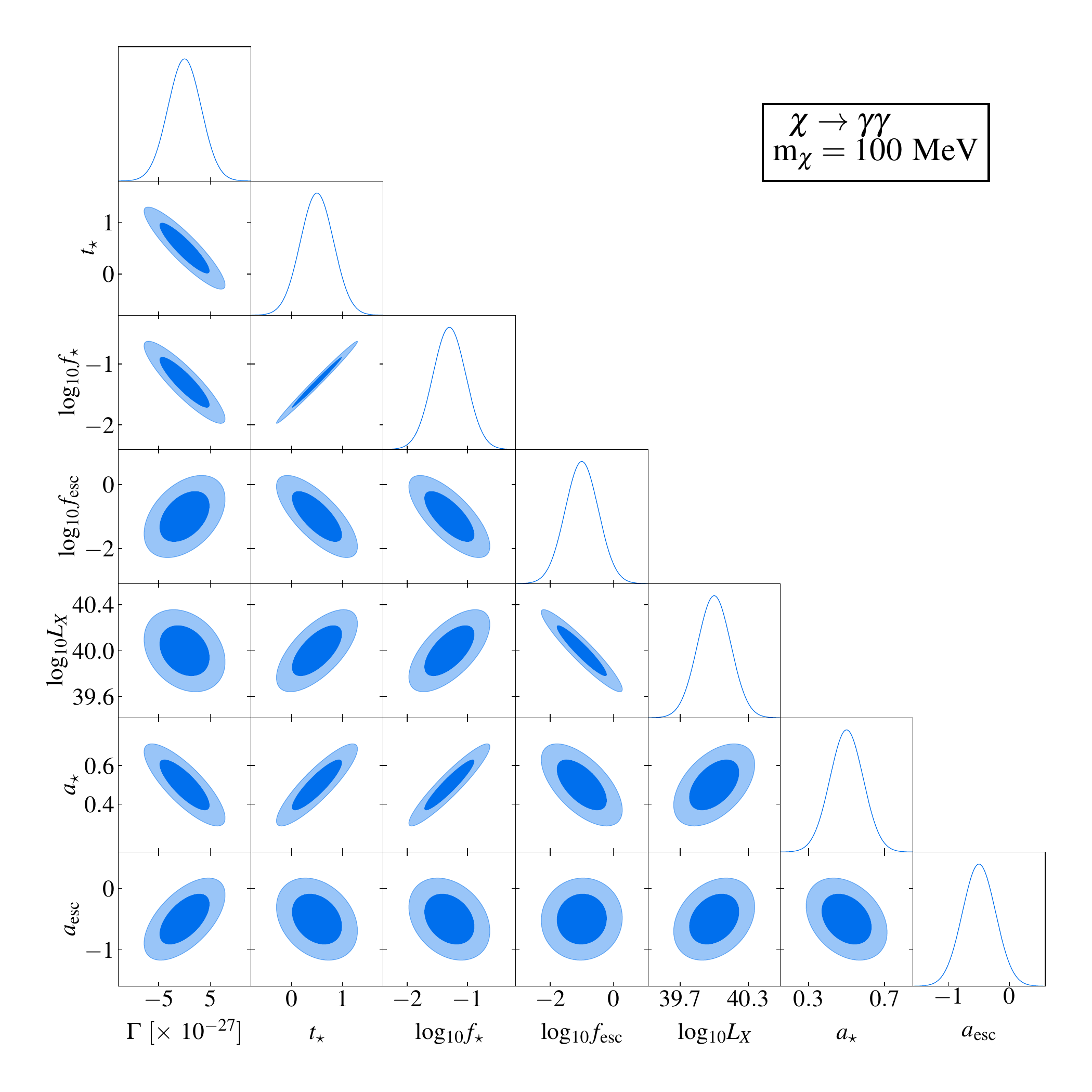}
\caption{Same as figure \ref{fig:ann_corry} but for $100$\,MeV \ac{DM} particles decaying into photon pairs.
}
\label{fig:dec_corry}
\end{figure}

\begin{figure}
    \includegraphics[width=1.0\textwidth]{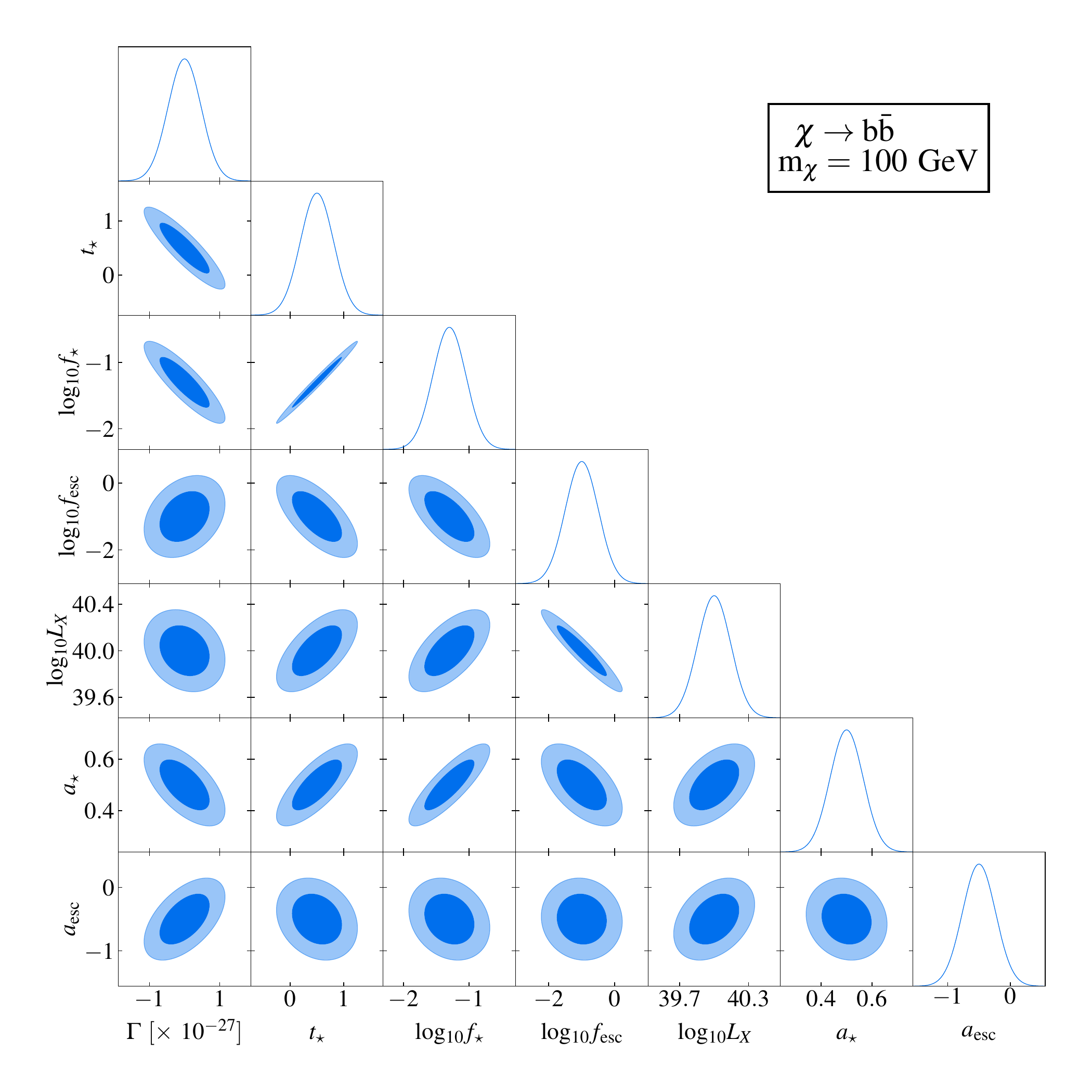}
    \caption{Same as figure \ref{fig:ann_corry} but for $100$\,GeV \ac{DM} particles decaying into bottom-anti-bottom quark pairs.}
    \label{fig:dec_corrb}
\end{figure}

{Remaining results of Fisher matrix forecast are shown in figures \ref{fig:ann_corry}--\ref{fig:dec_corrb}.
These figures demonstrate that, across all annihilation and decay channels, $\langle \sigma v \rangle$ and $\tau$ are positively correlated with $a_{\rm esc}$ and ${\rm log}_{10}f_{\rm esc}$, and negatively correlated with $t_{\star}$ and ${\rm log}_{10}f_{\star}$.
}

\acknowledgments

S.W. and X.Z. would like to express their gratitude to Xian Gao and Jiarui Sun for their warm hospitality during the final stage of this work, which was conducted during a visit to the Sun Yat-sen University. 
We are grateful to Kazunori Kohri, Yichao Li, Yue Shao, and Xukun Zhang for their valuable contributions to the discussion. 
This work is supported by the National SKA Program of China (Grant Nos. 2022SKA0110200, 2022SKA0110203), the National Key R\&D Program of China (Grant No. 2023YFC2206403), the National Natural Science Foundation of China (Grant Nos. 12175243, 12473001, and 11975072), and the National 111 Project (Grant No. B16009). 

\newpage

\bibliographystyle{JHEP}
\bibliography{biblio}

\providecommand{\href}[2]{#2}\begingroup\raggedright\begin{thebibliography}{10}

\bibitem{ParticleDataGroup:2024cfk}
{\scshape Particle Data Group} collaboration, \emph{{Review of particle
  physics}}, \href{https://doi.org/10.1103/PhysRevD.110.030001}{\emph{Phys.
  Rev. D} {\bfseries 110} (2024) 030001}.

\bibitem{Bechtol:2019acd}
K.~Bechtol et~al., \emph{{Dark Matter Science in the Era of LSST}},
  \href{https://arxiv.org/abs/1903.04425}{{\ttfamily 1903.04425}}.

\bibitem{Schumann:2019eaa}
M.~Schumann, \emph{{Direct Detection of WIMP Dark Matter: Concepts and
  Status}}, \href{https://doi.org/10.1088/1361-6471/ab2ea5}{\emph{J. Phys. G}
  {\bfseries 46} (2019) 103003}
  [\href{https://arxiv.org/abs/1903.03026}{{\ttfamily 1903.03026}}].

\bibitem{Bertone:2016nfn}
G.~Bertone and D.~Hooper, \emph{{History of dark matter}},
  \href{https://doi.org/10.1103/RevModPhys.90.045002}{\emph{Rev. Mod. Phys.}
  {\bfseries 90} (2018) 045002}
  [\href{https://arxiv.org/abs/1605.04909}{{\ttfamily 1605.04909}}].

\bibitem{Gaskins:2016cha}
J.M.~Gaskins, \emph{{A review of indirect searches for particle dark matter}},
  \href{https://doi.org/10.1080/00107514.2016.1175160}{\emph{Contemp. Phys.}
  {\bfseries 57} (2016) 496}
  [\href{https://arxiv.org/abs/1604.00014}{{\ttfamily 1604.00014}}].

\bibitem{Feng:2010gw}
J.L.~Feng, \emph{{Dark Matter Candidates from Particle Physics and Methods of
  Detection}},
  \href{https://doi.org/10.1146/annurev-astro-082708-101659}{\emph{Ann. Rev.
  Astron. Astrophys.} {\bfseries 48} (2010) 495}
  [\href{https://arxiv.org/abs/1003.0904}{{\ttfamily 1003.0904}}].

\bibitem{Bertone:2004pz}
G.~Bertone, D.~Hooper and J.~Silk, \emph{{Particle dark matter: Evidence,
  candidates and constraints}},
  \href{https://doi.org/10.1016/j.physrep.2004.08.031}{\emph{Phys. Rept.}
  {\bfseries 405} (2005) 279}
  [\href{https://arxiv.org/abs/hep-ph/0404175}{{\ttfamily hep-ph/0404175}}].

\bibitem{Sahni:2004ai}
V.~Sahni, \emph{{Dark matter and dark energy}},
  \href{https://doi.org/10.1007/b99562}{\emph{Lect. Notes Phys.} {\bfseries
  653} (2004) 141} [\href{https://arxiv.org/abs/astro-ph/0403324}{{\ttfamily
  astro-ph/0403324}}].

\bibitem{Smith:1988kw}
P.F.~Smith and J.D.~Lewin, \emph{{Dark Matter Detection}},
  \href{https://doi.org/10.1016/0370-1573(90)90081-C}{\emph{Phys. Rept.}
  {\bfseries 187} (1990) 203}.

\bibitem{Lin:2019uvt}
T.~Lin, \emph{{Dark matter models and direct detection}},
  \href{https://doi.org/10.22323/1.333.0009}{\emph{PoS} {\bfseries 333} (2019)
  009} [\href{https://arxiv.org/abs/1904.07915}{{\ttfamily 1904.07915}}].

\bibitem{Fermi-LAT:2016uux}
{\scshape Fermi-LAT, DES} collaboration, \emph{{Searching for Dark Matter
  Annihilation in Recently Discovered Milky Way Satellites with Fermi-LAT}},
  \href{https://doi.org/10.3847/1538-4357/834/2/110}{\emph{Astrophys. J.}
  {\bfseries 834} (2017) 110}
  [\href{https://arxiv.org/abs/1611.03184}{{\ttfamily 1611.03184}}].

\bibitem{MAGIC:2016xys}
{\scshape MAGIC, Fermi-LAT} collaboration, \emph{{Limits to Dark Matter
  Annihilation Cross-Section from a Combined Analysis of MAGIC and Fermi-LAT
  Observations of Dwarf Satellite Galaxies}},
  \href{https://doi.org/10.1088/1475-7516/2016/02/039}{\emph{JCAP} {\bfseries
  02} (2016) 039} [\href{https://arxiv.org/abs/1601.06590}{{\ttfamily
  1601.06590}}].

\bibitem{Slatyer:2016qyl}
T.R.~Slatyer and C.-L.~Wu, \emph{{General Constraints on Dark Matter Decay from
  the Cosmic Microwave Background}},
  \href{https://doi.org/10.1103/PhysRevD.95.023010}{\emph{Phys. Rev. D}
  {\bfseries 95} (2017) 023010}
  [\href{https://arxiv.org/abs/1610.06933}{{\ttfamily 1610.06933}}].

\bibitem{Fermi-LAT:2015att}
{\scshape Fermi-LAT} collaboration, \emph{{Searching for Dark Matter
  Annihilation from Milky Way Dwarf Spheroidal Galaxies with Six Years of Fermi
  Large Area Telescope Data}},
  \href{https://doi.org/10.1103/PhysRevLett.115.231301}{\emph{Phys. Rev. Lett.}
  {\bfseries 115} (2015) 231301}
  [\href{https://arxiv.org/abs/1503.02641}{{\ttfamily 1503.02641}}].

\bibitem{Slatyer:2015jla}
T.R.~Slatyer, \emph{{Indirect dark matter signatures in the cosmic dark ages.
  I. Generalizing the bound on s-wave dark matter annihilation from Planck
  results}}, \href{https://doi.org/10.1103/PhysRevD.93.023527}{\emph{Phys. Rev.
  D} {\bfseries 93} (2016) 023527}
  [\href{https://arxiv.org/abs/1506.03811}{{\ttfamily 1506.03811}}].

\bibitem{HESS:2013rld}
{\scshape H.E.S.S.} collaboration, \emph{{Search for Photon-Linelike Signatures
  from Dark Matter Annihilations with H.E.S.S.}},
  \href{https://doi.org/10.1103/PhysRevLett.110.041301}{\emph{Phys. Rev. Lett.}
  {\bfseries 110} (2013) 041301}
  [\href{https://arxiv.org/abs/1301.1173}{{\ttfamily 1301.1173}}].

\bibitem{Steigman:2012nb}
G.~Steigman, B.~Dasgupta and J.F.~Beacom, \emph{{Precise Relic WIMP Abundance
  and its Impact on Searches for Dark Matter Annihilation}},
  \href{https://doi.org/10.1103/PhysRevD.86.023506}{\emph{Phys. Rev. D}
  {\bfseries 86} (2012) 023506}
  [\href{https://arxiv.org/abs/1204.3622}{{\ttfamily 1204.3622}}].

\bibitem{Chen:2003gz}
X.-L.~Chen and M.~Kamionkowski, \emph{{Particle decays during the cosmic dark
  ages}}, \href{https://doi.org/10.1103/PhysRevD.70.043502}{\emph{Phys. Rev. D}
  {\bfseries 70} (2004) 043502}
  [\href{https://arxiv.org/abs/astro-ph/0310473}{{\ttfamily
  astro-ph/0310473}}].

\bibitem{Gondolo:1999ef}
P.~Gondolo and J.~Silk, \emph{{Dark matter annihilation at the galactic
  center}}, \href{https://doi.org/10.1103/PhysRevLett.83.1719}{\emph{Phys. Rev.
  Lett.} {\bfseries 83} (1999) 1719}
  [\href{https://arxiv.org/abs/astro-ph/9906391}{{\ttfamily
  astro-ph/9906391}}].

\bibitem{Cheung:2018vww}
K.~Cheung, J.-L.~Kuo, K.-W.~Ng and Y.-L.S.~Tsai, \emph{{The impact of EDGES
  21-cm data on dark matter interactions}},
  \href{https://doi.org/10.1016/j.physletb.2018.11.058}{\emph{Phys. Lett. B}
  {\bfseries 789} (2019) 137}
  [\href{https://arxiv.org/abs/1803.09398}{{\ttfamily 1803.09398}}].

\bibitem{Hawking:1971ei}
S.~Hawking, \emph{{Gravitationally collapsed objects of very low mass}},
  \href{https://doi.org/10.1093/mnras/152.1.75}{\emph{Mon. Not. Roy. Astron.
  Soc.} {\bfseries 152} (1971) 75}.

\bibitem{Sasaki:2018dmp}
M.~Sasaki, T.~Suyama, T.~Tanaka and S.~Yokoyama, \emph{{Primordial black
  holes\textemdash{}perspectives in gravitational wave astronomy}},
  \href{https://doi.org/10.1088/1361-6382/aaa7b4}{\emph{Class. Quant. Grav.}
  {\bfseries 35} (2018) 063001}
  [\href{https://arxiv.org/abs/1801.05235}{{\ttfamily 1801.05235}}].

\bibitem{Carr:2020gox}
B.~Carr, K.~Kohri, Y.~Sendouda and J.~Yokoyama, \emph{{Constraints on
  primordial black holes}},
  \href{https://doi.org/10.1088/1361-6633/ac1e31}{\emph{Rept. Prog. Phys.}
  {\bfseries 84} (2021) 116902}
  [\href{https://arxiv.org/abs/2002.12778}{{\ttfamily 2002.12778}}].

\bibitem{Green:2020jor}
A.M.~Green and B.J.~Kavanagh, \emph{{Primordial Black Holes as a dark matter
  candidate}}, \href{https://doi.org/10.1088/1361-6471/abc534}{\emph{J. Phys.
  G} {\bfseries 48} (2021) 043001}
  [\href{https://arxiv.org/abs/2007.10722}{{\ttfamily 2007.10722}}].

\bibitem{Carr:2021bzv}
B.~Carr and F.~Kuhnel, \emph{{Primordial black holes as dark matter
  candidates}},
  \href{https://doi.org/10.21468/SciPostPhysLectNotes.48}{\emph{SciPost Phys.
  Lect. Notes} {\bfseries 48} (2022) 1}
  [\href{https://arxiv.org/abs/2110.02821}{{\ttfamily 2110.02821}}].

\bibitem{Auffinger:2022ive}
J.~Auffinger, \emph{{Primordial black holes as dark matter and Hawking
  radiation constraints with BlackHawk}}, Ph.D. thesis, Institut de Physique
  des 2 Infinis de Lyon, France, IP2I, Lyon, 2022.

\bibitem{Mittal:2021egv}
S.~Mittal, A.~Ray, G.~Kulkarni and B.~Dasgupta, \emph{{Constraining primordial
  black holes as dark matter using the global 21-cm signal with X-ray heating
  and excess radio background}},
  \href{https://doi.org/10.1088/1475-7516/2022/03/030}{\emph{JCAP} {\bfseries
  03} (2022) 030} [\href{https://arxiv.org/abs/2107.02190}{{\ttfamily
  2107.02190}}].

\bibitem{Mena:2019nhm}
O.~Mena, S.~Palomares-Ruiz, P.~Villanueva-Domingo and S.J.~Witte,
  \emph{{Constraining the primordial black hole abundance with 21-cm
  cosmology}}, \href{https://doi.org/10.1103/PhysRevD.100.043540}{\emph{Phys.
  Rev. D} {\bfseries 100} (2019) 043540}
  [\href{https://arxiv.org/abs/1906.07735}{{\ttfamily 1906.07735}}].

\bibitem{Saha:2021pqf}
A.K.~Saha and R.~Laha, \emph{{Sensitivities on nonspinning and spinning
  primordial black hole dark matter with global 21-cm troughs}},
  \href{https://doi.org/10.1103/PhysRevD.105.103026}{\emph{Phys. Rev. D}
  {\bfseries 105} (2022) 103026}
  [\href{https://arxiv.org/abs/2112.10794}{{\ttfamily 2112.10794}}].

\bibitem{Facchinetti:2023slb}
G.~Facchinetti, L.~Lopez-Honorez, Y.~Qin and A.~Mesinger, \emph{{21cm signal
  sensitivity to dark matter decay}},
  \href{https://doi.org/10.1088/1475-7516/2024/01/005}{\emph{JCAP} {\bfseries
  01} (2024) 005} [\href{https://arxiv.org/abs/2308.16656}{{\ttfamily
  2308.16656}}].

\bibitem{Shao:2023agv}
Y.~Shao, Y.~Xu, Y.~Wang, W.~Yang, R.~Li, X.~Zhang et~al., \emph{{The 21-cm
  forest as a simultaneous probe of dark matter and cosmic heating history}},
  \href{https://doi.org/10.1038/s41550-023-02024-7}{\emph{Nature Astron.}
  {\bfseries 7} (2023) 1116}
  [\href{https://arxiv.org/abs/2307.04130}{{\ttfamily 2307.04130}}].

\bibitem{Novosyadlyj:2024bie}
B.~Novosyadlyj, Y.~Kulinich and D.~Koval, \emph{{Global signal in the
  redshifted hydrogen 21-cm line from Dark Ages and Cosmic Dawn: dependences on
  dark matter nature and first light}},
  \href{https://arxiv.org/abs/2410.07380}{{\ttfamily 2410.07380}}.

\bibitem{Sun:2023acy}
Y.~Sun, J.W.~Foster, H.~Liu, J.B.~Mu\~noz and T.R.~Slatyer,
  \emph{{Inhomogeneous Energy Injection in the 21-cm Power Spectrum:
  Sensitivity to Dark Matter Decay}},
  \href{https://arxiv.org/abs/2312.11608}{{\ttfamily 2312.11608}}.

\bibitem{Liu:2023fgu}
H.~Liu, W.~Qin, G.W.~Ridgway and T.R.~Slatyer, \emph{{Exotic energy injection
  in the early Universe. I. A novel treatment for low-energy electrons and
  photons}}, \href{https://doi.org/10.1103/PhysRevD.108.043530}{\emph{Phys.
  Rev. D} {\bfseries 108} (2023) 043530}
  [\href{https://arxiv.org/abs/2303.07366}{{\ttfamily 2303.07366}}].

\bibitem{Liu:2018uzy}
H.~Liu and T.R.~Slatyer, \emph{{Implications of a 21-cm signal for dark matter
  annihilation and decay}},
  \href{https://doi.org/10.1103/PhysRevD.98.023501}{\emph{Phys. Rev. D}
  {\bfseries 98} (2018) 023501}
  [\href{https://arxiv.org/abs/1803.09739}{{\ttfamily 1803.09739}}].

\bibitem{Liu:2023nct}
H.~Liu, W.~Qin, G.W.~Ridgway and T.R.~Slatyer, \emph{{Exotic energy injection
  in the early Universe. II. CMB spectral distortions and constraints on light
  dark matter}}, \href{https://doi.org/10.1103/PhysRevD.108.043531}{\emph{Phys.
  Rev. D} {\bfseries 108} (2023) 043531}
  [\href{https://arxiv.org/abs/2303.07370}{{\ttfamily 2303.07370}}].

\bibitem{Xu:2024vdn}
C.~Xu, W.~Qin and T.R.~Slatyer, \emph{{CMB limits on decaying dark matter
  beyond the ionization threshold}},
  \href{https://doi.org/10.1103/PhysRevD.110.123529}{\emph{Phys. Rev. D}
  {\bfseries 110} (2024) 123529}
  [\href{https://arxiv.org/abs/2408.13305}{{\ttfamily 2408.13305}}].

\bibitem{Capozzi:2024gqy}
F.~Capozzi, R.Z.~Ferreira, L.~Lopez-Honorez and O.~Mena, \emph{{CMB and
  Lyman-$\alpha$ constraints on dark matter decays to photons}},  in
  \emph{{Beyond Standard Model: From Theory to Experiment}}, 2024,
  \href{https://doi.org/10.31526/ACP.BSM-2023.2}{DOI}.

\bibitem{Liu:2020wqz}
H.~Liu, W.~Qin, G.W.~Ridgway and T.R.~Slatyer, \emph{{Lyman-\ensuremath{\alpha}
  constraints on cosmic heating from dark matter annihilation and decay}},
  \href{https://doi.org/10.1103/PhysRevD.104.043514}{\emph{Phys. Rev. D}
  {\bfseries 104} (2021) 043514}
  [\href{https://arxiv.org/abs/2008.01084}{{\ttfamily 2008.01084}}].

\bibitem{Bowman:2018yin}
J.D.~Bowman, A.E.E.~Rogers, R.A.~Monsalve, T.J.~Mozdzen and N.~Mahesh,
  \emph{{An absorption profile centred at 78 megahertz in the sky-averaged
  spectrum}}, \href{https://doi.org/10.1038/nature25792}{\emph{Nature}
  {\bfseries 555} (2018) 67}
  [\href{https://arxiv.org/abs/1810.05912}{{\ttfamily 1810.05912}}].

\bibitem{Cang:2021owu}
J.~Cang, Y.~Gao and Y.-Z.~Ma, \emph{{21-cm constraints on spinning primordial
  black holes}},
  \href{https://doi.org/10.1088/1475-7516/2022/03/012}{\emph{JCAP} {\bfseries
  03} (2022) 012} [\href{https://arxiv.org/abs/2108.13256}{{\ttfamily
  2108.13256}}].

\bibitem{Bevins:2022ajf}
H.T.J.~Bevins, A.~Fialkov, E.d.L.~Acedo, W.J.~Handley, S.~Singh,
  R.~Subrahmanyan et~al., \emph{{Astrophysical constraints from the SARAS 3
  non-detection of the cosmic dawn sky-averaged 21-cm signal}},
  \href{https://doi.org/10.1038/s41550-022-01825-6}{\emph{Nature Astron.}
  {\bfseries 6} (2022) 1473}
  [\href{https://arxiv.org/abs/2212.00464}{{\ttfamily 2212.00464}}].

\bibitem{2023ChJSS..43...43C}
X.~{Chen}, J.~{Yan}, Y.~{Xu}, L.~{Deng}, F.~{Wu}, L.~{Wu} et~al.,
  \emph{{Discovering the Sky at the Longest Wavelength Mission - A Pathfinder
  for Exploring the Cosmic Dark Ages.}},
  \href{https://doi.org/10.11728/cjss2023.01.220104001}{\emph{Chinese Journal
  of Space Science} {\bfseries 43} (2023) 43}.

\bibitem{Planck:2018vyg}
{\scshape Planck} collaboration, \emph{{Planck 2018 results. VI. Cosmological
  parameters}},
  \href{https://doi.org/10.1051/0004-6361/201833910}{\emph{Astron. Astrophys.}
  {\bfseries 641} (2020) A6}
  [\href{https://arxiv.org/abs/1807.06209}{{\ttfamily 1807.06209}}].

\bibitem{Slatyer:2015kla}
T.R.~Slatyer, \emph{{Indirect Dark Matter Signatures in the Cosmic Dark Ages
  II. Ionization, Heating and Photon Production from Arbitrary Energy
  Injections}}, \href{https://doi.org/10.1103/PhysRevD.93.023521}{\emph{Phys.
  Rev. D} {\bfseries 93} (2016) 023521}
  [\href{https://arxiv.org/abs/1506.03812}{{\ttfamily 1506.03812}}].

\bibitem{Cirelli:2010xx}
M.~Cirelli, G.~Corcella, A.~Hektor, G.~Hutsi, M.~Kadastik, P.~Panci et~al.,
  \emph{{PPPC 4 DM ID: A Poor Particle Physicist Cookbook for Dark Matter
  Indirect Detection}},
  \href{https://doi.org/10.1088/1475-7516/2012/10/E01}{\emph{JCAP} {\bfseries
  03} (2011) 051} [\href{https://arxiv.org/abs/1012.4515}{{\ttfamily
  1012.4515}}].

\bibitem{Bierlich:2022pfr}
C.~Bierlich et~al., \emph{{A comprehensive guide to the physics and usage of
  PYTHIA 8.3}},
  \href{https://doi.org/10.21468/SciPostPhysCodeb.8}{\emph{SciPost Phys.
  Codeb.} {\bfseries 2022} (2022) 8}
  [\href{https://arxiv.org/abs/2203.11601}{{\ttfamily 2203.11601}}].

\bibitem{Slatyer:2012yq}
T.R.~Slatyer, \emph{{Energy Injection And Absorption In The Cosmic Dark Ages}},
  \href{https://doi.org/10.1103/PhysRevD.87.123513}{\emph{Phys. Rev. D}
  {\bfseries 87} (2013) 123513}
  [\href{https://arxiv.org/abs/1211.0283}{{\ttfamily 1211.0283}}].

\bibitem{Slatyer:2009yq}
T.R.~Slatyer, N.~Padmanabhan and D.P.~Finkbeiner, \emph{{CMB Constraints on
  WIMP Annihilation: Energy Absorption During the Recombination Epoch}},
  \href{https://doi.org/10.1103/PhysRevD.80.043526}{\emph{Phys. Rev. D}
  {\bfseries 80} (2009) 043526}
  [\href{https://arxiv.org/abs/0906.1197}{{\ttfamily 0906.1197}}].

\bibitem{Liu:2019bbm}
H.~Liu, G.W.~Ridgway and T.R.~Slatyer, \emph{{Code package for calculating
  modified cosmic ionization and thermal histories with dark matter and other
  exotic energy injections}},
  \href{https://doi.org/10.1103/PhysRevD.101.023530}{\emph{Phys. Rev. D}
  {\bfseries 101} (2020) 023530}
  [\href{https://arxiv.org/abs/1904.09296}{{\ttfamily 1904.09296}}].

\bibitem{Takahashi:2021pse}
R.~Takahashi and K.~Kohri, \emph{{Cosmological boost factor for dark matter
  annihilation at redshifts of z=10-100 using the power spectrum approach}},
  \href{https://doi.org/10.1103/PhysRevD.104.103518}{\emph{Phys. Rev. D}
  {\bfseries 104} (2021) 103518}
  [\href{https://arxiv.org/abs/2107.00897}{{\ttfamily 2107.00897}}].

\bibitem{Auffinger:2020ztk}
J.~Auffinger and A.~Arbey, \emph{{BlackHawk: A tool for computing Black Hole
  evaporation}}, \href{https://doi.org/10.22323/1.392.0024}{\emph{PoS}
  {\bfseries TOOLS2020} (2021) 024}
  [\href{https://arxiv.org/abs/2012.12902}{{\ttfamily 2012.12902}}].

\bibitem{Poulin:2017bwe}
V.~Poulin, P.D.~Serpico, F.~Calore, S.~Clesse and K.~Kohri, \emph{{CMB bounds
  on disk-accreting massive primordial black holes}},
  \href{https://doi.org/10.1103/PhysRevD.96.083524}{\emph{Phys. Rev. D}
  {\bfseries 96} (2017) 083524}
  [\href{https://arxiv.org/abs/1707.04206}{{\ttfamily 1707.04206}}].

\bibitem{Furlanetto:2006jb}
S.~Furlanetto, S.P.~Oh and F.~Briggs, \emph{{Cosmology at Low Frequencies: The
  21 cm Transition and the High-Redshift Universe}},
  \href{https://doi.org/10.1016/j.physrep.2006.08.002}{\emph{Phys. Rept.}
  {\bfseries 433} (2006) 181}
  [\href{https://arxiv.org/abs/astro-ph/0608032}{{\ttfamily
  astro-ph/0608032}}].

\bibitem{Pritchard:2011xb}
J.R.~Pritchard and A.~Loeb, \emph{{21-cm cosmology}},
  \href{https://doi.org/10.1088/0034-4885/75/8/086901}{\emph{Rept. Prog. Phys.}
  {\bfseries 75} (2012) 086901}
  [\href{https://arxiv.org/abs/1109.6012}{{\ttfamily 1109.6012}}].

\bibitem{Mesinger:2010ne}
A.~Mesinger, S.~Furlanetto and R.~Cen, \emph{{21cmFAST: A Fast, Semi-Numerical
  Simulation of the High-Redshift 21-cm Signal}},
  \href{https://doi.org/10.1111/j.1365-2966.2010.17731.x}{\emph{Mon. Not. Roy.
  Astron. Soc.} {\bfseries 411} (2011) 955}
  [\href{https://arxiv.org/abs/1003.3878}{{\ttfamily 1003.3878}}].

\bibitem{Tegmark:1996bz}
M.~Tegmark, A.~Taylor and A.~Heavens, \emph{{Karhunen-Loeve eigenvalue problems
  in cosmology: How should we tackle large data sets?}},
  \href{https://doi.org/10.1086/303939}{\emph{Astrophys. J.} {\bfseries 480}
  (1997) 22} [\href{https://arxiv.org/abs/astro-ph/9603021}{{\ttfamily
  astro-ph/9603021}}].

\bibitem{Pritchard:2010pa}
J.R.~Pritchard and A.~Loeb, \emph{{Constraining the unexplored period between
  the dark ages and reionization with observations of the global 21 cm
  signal}}, \href{https://doi.org/10.1103/PhysRevD.82.023006}{\emph{Phys. Rev.
  D} {\bfseries 82} (2010) 023006}
  [\href{https://arxiv.org/abs/1005.4057}{{\ttfamily 1005.4057}}].

\bibitem{Liu:2019ygl}
X.-W.~Liu, C.~Heneka and L.~Amendola, \emph{{Constraining coupled quintessence
  with the 21cm signal}},
  \href{https://doi.org/10.1088/1475-7516/2020/05/038}{\emph{JCAP} {\bfseries
  05} (2020) 038} [\href{https://arxiv.org/abs/1910.02763}{{\ttfamily
  1910.02763}}].

\bibitem{Jester:2009dw}
S.~Jester and H.~Falcke, \emph{{Science with a lunar low-frequency array: from
  the dark ages of the Universe to nearby exoplanets}},
  \href{https://doi.org/10.1016/j.newar.2009.02.001}{\emph{New Astron. Rev.}
  {\bfseries 53} (2009) 1} [\href{https://arxiv.org/abs/0902.0493}{{\ttfamily
  0902.0493}}].

\bibitem{deOliveira-Costa:2008cxd}
A.~de~Oliveira-Costa, M.~Tegmark, B.M.~Gaensler, J.~Jonas, T.L.~Landecker and
  P.~Reich, \emph{{A model of diffuse Galactic Radio Emission from 10 MHz to
  100 GHz}}, \href{https://doi.org/10.1111/j.1365-2966.2008.13376.x}{\emph{Mon.
  Not. Roy. Astron. Soc.} {\bfseries 388} (2008) 247}
  [\href{https://arxiv.org/abs/0802.1525}{{\ttfamily 0802.1525}}].

\bibitem{Zhang:2023usm}
Z.-X.~Zhang, Y.-M.~Wang, J.~Cang, Z.~Zhang, Y.~Liu, S.-Y.~Li et~al.,
  \emph{{Dark matter search with CMB: a~study of foregrounds}},
  \href{https://doi.org/10.1088/1475-7516/2023/10/002}{\emph{JCAP} {\bfseries
  10} (2023) 002} [\href{https://arxiv.org/abs/2304.07793}{{\ttfamily
  2304.07793}}].

\bibitem{Cirelli:2020bpc}
M.~Cirelli, N.~Fornengo, B.J.~Kavanagh and E.~Pinetti, \emph{{Integral X-ray
  constraints on sub-GeV Dark Matter}},
  \href{https://doi.org/10.1103/PhysRevD.103.063022}{\emph{Phys. Rev. D}
  {\bfseries 103} (2021) 063022}
  [\href{https://arxiv.org/abs/2007.11493}{{\ttfamily 2007.11493}}].

\bibitem{HESS:2018kom}
{\scshape HESS} collaboration, \emph{{Searches for gamma-ray lines and 'pure
  WIMP' spectra from Dark Matter annihilations in dwarf galaxies with
  H.E.S.S}}, \href{https://doi.org/10.1088/1475-7516/2018/11/037}{\emph{JCAP}
  {\bfseries 11} (2018) 037}
  [\href{https://arxiv.org/abs/1810.00995}{{\ttfamily 1810.00995}}].

\bibitem{HESS:2014zqa}
{\scshape H.E.S.S.} collaboration, \emph{{Search for dark matter annihilation
  signatures in H.E.S.S. observations of Dwarf Spheroidal Galaxies}},
  \href{https://doi.org/10.1103/PhysRevD.90.112012}{\emph{Phys. Rev. D}
  {\bfseries 90} (2014) 112012}
  [\href{https://arxiv.org/abs/1410.2589}{{\ttfamily 1410.2589}}].

\bibitem{VERITAS:2017tif}
{\scshape VERITAS} collaboration, \emph{{Dark Matter Constraints from a Joint
  Analysis of Dwarf Spheroidal Galaxy Observations with VERITAS}},
  \href{https://doi.org/10.1103/PhysRevD.95.082001}{\emph{Phys. Rev. D}
  {\bfseries 95} (2017) 082001}
  [\href{https://arxiv.org/abs/1703.04937}{{\ttfamily 1703.04937}}].

\bibitem{MAGIC:2017avy}
{\scshape MAGIC} collaboration, \emph{{Indirect dark matter searches in the
  dwarf satellite galaxy Ursa Major II with the MAGIC Telescopes}},
  \href{https://doi.org/10.1088/1475-7516/2018/03/009}{\emph{JCAP} {\bfseries
  03} (2018) 009} [\href{https://arxiv.org/abs/1712.03095}{{\ttfamily
  1712.03095}}].

\bibitem{Aleksic:2013xea}
J.~Aleksi\'c et~al., \emph{{Optimized dark matter searches in deep observations
  of Segue 1 with MAGIC}},
  \href{https://doi.org/10.1088/1475-7516/2014/02/008}{\emph{JCAP} {\bfseries
  02} (2014) 008} [\href{https://arxiv.org/abs/1312.1535}{{\ttfamily
  1312.1535}}].

\bibitem{Boudaud:2016mos}
M.~Boudaud, J.~Lavalle and P.~Salati, \emph{{Novel cosmic-ray electron and
  positron constraints on MeV dark matter particles}},
  \href{https://doi.org/10.1103/PhysRevLett.119.021103}{\emph{Phys. Rev. Lett.}
  {\bfseries 119} (2017) 021103}
  [\href{https://arxiv.org/abs/1612.07698}{{\ttfamily 1612.07698}}].

\bibitem{Boudaud:2018oya}
M.~Boudaud, T.~Lacroix, M.~Stref and J.~Lavalle, \emph{{Robust cosmic-ray
  constraints on $p$-wave annihilating MeV dark matter}},
  \href{https://doi.org/10.1103/PhysRevD.99.061302}{\emph{Phys. Rev. D}
  {\bfseries 99} (2019) 061302}
  [\href{https://arxiv.org/abs/1810.01680}{{\ttfamily 1810.01680}}].

\bibitem{Cohen:2016uyg}
T.~Cohen, K.~Murase, N.L.~Rodd, B.R.~Safdi and Y.~Soreq,
  \emph{{\ensuremath{\gamma} -ray Constraints on Decaying Dark Matter and
  Implications for IceCube}},
  \href{https://doi.org/10.1103/PhysRevLett.119.021102}{\emph{Phys. Rev. Lett.}
  {\bfseries 119} (2017) 021102}
  [\href{https://arxiv.org/abs/1612.05638}{{\ttfamily 1612.05638}}].

\bibitem{Boudaud:2018hqb}
M.~Boudaud and M.~Cirelli, \emph{{Voyager 1 $e^\pm$ Further Constrain
  Primordial Black Holes as Dark Matter}},
  \href{https://doi.org/10.1103/PhysRevLett.122.041104}{\emph{Phys. Rev. Lett.}
  {\bfseries 122} (2019) 041104}
  [\href{https://arxiv.org/abs/1807.03075}{{\ttfamily 1807.03075}}].

\bibitem{Capozzi:2023xie}
F.~Capozzi, R.Z.~Ferreira, L.~Lopez-Honorez and O.~Mena, \emph{{CMB and
  Lyman-\ensuremath{\alpha} constraints on dark matter decays to photons}},
  \href{https://doi.org/10.1088/1475-7516/2023/06/060}{\emph{JCAP} {\bfseries
  06} (2023) 060} [\href{https://arxiv.org/abs/2303.07426}{{\ttfamily
  2303.07426}}].

\bibitem{Essig:2013goa}
R.~Essig, E.~Kuflik, S.D.~McDermott, T.~Volansky and K.M.~Zurek,
  \emph{{Constraining Light Dark Matter with Diffuse X-Ray and Gamma-Ray
  Observations}}, \href{https://doi.org/10.1007/JHEP11(2013)193}{\emph{JHEP}
  {\bfseries 11} (2013) 193} [\href{https://arxiv.org/abs/1309.4091}{{\ttfamily
  1309.4091}}].

\bibitem{Massari:2015xea}
A.~Massari, E.~Izaguirre, R.~Essig, A.~Albert, E.~Bloom and
  G.A.~G\'omez-Vargas, \emph{{Strong Optimized Conservative $Fermi$-LAT
  Constraints on Dark Matter Models from the Inclusive Photon Spectrum}},
  \href{https://doi.org/10.1103/PhysRevD.91.083539}{\emph{Phys. Rev. D}
  {\bfseries 91} (2015) 083539}
  [\href{https://arxiv.org/abs/1503.07169}{{\ttfamily 1503.07169}}].

\bibitem{Cadamuro:2011fd}
D.~Cadamuro and J.~Redondo, \emph{{Cosmological bounds on pseudo
  Nambu-Goldstone bosons}},
  \href{https://doi.org/10.1088/1475-7516/2012/02/032}{\emph{JCAP} {\bfseries
  02} (2012) 032} [\href{https://arxiv.org/abs/1110.2895}{{\ttfamily
  1110.2895}}].

\bibitem{Koechler:2023ual}
J.~Koechler, \emph{{X-rays constraints on sub-GeV Dark Matter}},  in \emph{{TeV
  Particle Astrophysics 2023}}, 9, 2023
  [\href{https://arxiv.org/abs/2309.10043}{{\ttfamily 2309.10043}}].

\bibitem{Calore:2022pks}
F.~Calore, A.~Dekker, P.D.~Serpico and T.~Siegert, \emph{{Constraints on light
  decaying dark matter candidates from 16~yr of INTEGRAL/SPI observations}},
  \href{https://doi.org/10.1093/mnras/stad457}{\emph{Mon. Not. Roy. Astron.
  Soc.} {\bfseries 520} (2023) 4167}
  [\href{https://arxiv.org/abs/2209.06299}{{\ttfamily 2209.06299}}].

\bibitem{Foster:2022nva}
J.W.~Foster, Y.~Park, B.R.~Safdi, Y.~Soreq and W.L.~Xu, \emph{{Search for dark
  matter lines at the Galactic Center with 14~years of Fermi data}},
  \href{https://doi.org/10.1103/PhysRevD.107.103047}{\emph{Phys. Rev. D}
  {\bfseries 107} (2023) 103047}
  [\href{https://arxiv.org/abs/2212.07435}{{\ttfamily 2212.07435}}].

\bibitem{Poulin:2016anj}
V.~Poulin, J.~Lesgourgues and P.D.~Serpico, \emph{{Cosmological constraints on
  exotic injection of electromagnetic energy}},
  \href{https://doi.org/10.1088/1475-7516/2017/03/043}{\emph{JCAP} {\bfseries
  03} (2017) 043} [\href{https://arxiv.org/abs/1610.10051}{{\ttfamily
  1610.10051}}].

\bibitem{Wang:2020uvi}
S.~Wang, D.-M.~Xia, X.~Zhang, S.~Zhou and Z.~Chang, \emph{{Constraining
  primordial black holes as dark matter at JUNO}},
  \href{https://doi.org/10.1103/PhysRevD.103.043010}{\emph{Phys. Rev. D}
  {\bfseries 103} (2021) 043010}
  [\href{https://arxiv.org/abs/2010.16053}{{\ttfamily 2010.16053}}].

\bibitem{Carr:2016hva}
B.J.~Carr, K.~Kohri, Y.~Sendouda and J.~Yokoyama, \emph{{Constraints on
  primordial black holes from the Galactic gamma-ray background}},
  \href{https://doi.org/10.1103/PhysRevD.94.044029}{\emph{Phys. Rev. D}
  {\bfseries 94} (2016) 044029}
  [\href{https://arxiv.org/abs/1604.05349}{{\ttfamily 1604.05349}}].

\bibitem{Chluba:2020oip}
J.~Chluba, A.~Ravenni and S.K.~Acharya, \emph{{Thermalization of large energy
  release in the early Universe}},
  \href{https://doi.org/10.1093/mnras/staa2131}{\emph{Mon. Not. Roy. Astron.
  Soc.} {\bfseries 498} (2020) 959}
  [\href{https://arxiv.org/abs/2005.11325}{{\ttfamily 2005.11325}}].

\bibitem{Acharya:2020jbv}
S.K.~Acharya and R.~Khatri, \emph{{CMB and BBN constraints on evaporating
  primordial black holes revisited}},
  \href{https://doi.org/10.1088/1475-7516/2020/06/018}{\emph{JCAP} {\bfseries
  06} (2020) 018} [\href{https://arxiv.org/abs/2002.00898}{{\ttfamily
  2002.00898}}].

\bibitem{Clark:2016nst}
S.~Clark, B.~Dutta, Y.~Gao, L.E.~Strigari and S.~Watson, \emph{{Planck
  Constraint on Relic Primordial Black Holes}},
  \href{https://doi.org/10.1103/PhysRevD.95.083006}{\emph{Phys. Rev. D}
  {\bfseries 95} (2017) 083006}
  [\href{https://arxiv.org/abs/1612.07738}{{\ttfamily 1612.07738}}].

\bibitem{Carr:2009jm}
B.J.~Carr, K.~Kohri, Y.~Sendouda and J.~Yokoyama, \emph{{New cosmological
  constraints on primordial black holes}},
  \href{https://doi.org/10.1103/PhysRevD.81.104019}{\emph{Phys. Rev. D}
  {\bfseries 81} (2010) 104019}
  [\href{https://arxiv.org/abs/0912.5297}{{\ttfamily 0912.5297}}].

\bibitem{Pattison:2023eej}
J.H.N.~Pattison, D.J.~Anstey and E.d.L.~Acedo, \emph{{Modelling a hot horizon
  in global 21-cm experimental foregrounds}},
  \href{https://doi.org/10.1093/mnras/stad3378}{\emph{Mon. Not. Roy. Astron.
  Soc.} {\bfseries 527} (2023) 2413}
  [\href{https://arxiv.org/abs/2307.02908}{{\ttfamily 2307.02908}}].

\bibitem{Liu:2019awk}
A.~Liu and J.R.~Shaw, \emph{{Data Analysis for Precision 21 cm Cosmology}},
  \href{https://doi.org/10.1088/1538-3873/ab5bfd}{\emph{Publ. Astron. Soc.
  Pac.} {\bfseries 132} (2020) 062001}
  [\href{https://arxiv.org/abs/1907.08211}{{\ttfamily 1907.08211}}].

\bibitem{Deshpande:2018ugh}
A.A.~Deshpande, \emph{{Dipole Anisotropy as an Essential Qualifier for the
  Monopole Component of the Cosmic-Dawn Spectral Signature, and the Potential
  of Diurnal Pattern for Foreground Estimation}},
  \href{https://doi.org/10.3847/2041-8213/aae318}{\emph{Astrophys. J. Lett.}
  {\bfseries 866} (2018) L7}
  [\href{https://arxiv.org/abs/1804.07993}{{\ttfamily 1804.07993}}].

\bibitem{Tripathi:2024iai}
A.~Tripathi, A.~Datta, M.~Choudhury and S.~Majumdar, \emph{{Extracting the
  Global 21-cm signal from Cosmic Dawn and Epoch of Reionization in the
  presence of Foreground and Ionosphere}},
  \href{https://doi.org/10.1093/mnras/stae078}{\emph{Mon. Not. Roy. Astron.
  Soc.} {\bfseries 528} (2024) 1945}
  [\href{https://arxiv.org/abs/2401.01935}{{\ttfamily 2401.01935}}].

\end{thebibliography}\endgroup

\end{document}